\documentclass[a4paper,fleqn,usenatbib]{mnras}

\usepackage[T1]{fontenc}
\usepackage{ae,aecompl}

\usepackage{graphicx}	% Including figure files
\usepackage{amsmath}	% Advanced maths commands
\usepackage{amssymb}	% Extra maths symbols
\usepackage[normalem]{ulem}

\newcommand\bcdot{{\bmath\cdot}}
\newcommand\btimes{{\bmath\times}}

\newcommand\grad{{\bmath\nabla}}

\renewcommand\div{{\bmath\nabla}{\bmath\cdot}}

\newcommand\bOmega{{\bmath\Omega}}

\newcommand\bmB{{\bmath B}}

\newcommand\bmf{{\bmath f}}

\newcommand\bmr{{\bmath r}}

\renewcommand\bmu{{\bmath u}}

\newcommand\rmc{\mathrm{c}}
\newcommand\rmd{\mathrm{d}}

\newcommand\rmg{\mathrm{g}}

\newcommand\rmw{\mathrm{w}}
\newcommand\rmnw{\mathrm{nw}}

\newcommand\f{\frac}

\title[Tidal dissipation: the presence of a magnetic field]{Tidal dissipation in rotating fluid bodies: the presence of a magnetic field}

\author[Y. Lin and G. I. Ogilvie]{
Yufeng Lin\thanks{E-mail: yl552@cam.ac.uk} and 
Gordon I. Ogilvie
\\
Department of Applied Mathematics and Theoretical Physics, University of Cambridge, Centre for Mathematical Sciences, \\
Wilberforce Road, Cambridge CB3 0WA, UK
}

\date{Accepted XXX. Received YYY; in original form ZZZ}

\pubyear{2017}

\begin{document}
\label{firstpage}
\pagerange{\pageref{firstpage}--\pageref{lastpage}}
\maketitle

\begin{abstract}
We investigate effects of the presence of a magnetic field on tidal dissipation in rotating fluid bodies. We consider a simplified model consisting of a rigid core and a fluid envelope, permeated by a background magnetic field (either a dipolar field or a uniform axial field). The wavelike tidal responses in the fluid layer are in the form of magnetic-Coriolis waves, which are restored by both the Coriolis force and the Lorentz force. Energy dissipation occurs through viscous damping and Ohmic damping of these waves. Our numerical results show that the tidal dissipation can be dominated by Ohmic damping even with a weak magnetic field. The presence of a magnetic field smooths out the complicated frequency-dependence of the dissipation rate, and broadens the frequency spectrum of the dissipation rate, depending on the strength of the background magnetic field. However, the frequency-averaged dissipation is independent of the strength and structure of the magnetic field, and of the dissipative parameters, in the approximation that the wave-like response is driven only by the Coriolis force acting on the non-wavelike tidal flow. Indeed, the frequency-averaged dissipation quantity is in good agreement with previous analytical results in the absence of magnetic fields. Our results suggest that the frequency-averaged tidal dissipation of the wavelike perturbations is insensitive to detailed damping mechanisms and dissipative properties.  
\end{abstract}


%%%%%%%%%%%%%%%%%%%%%%%%%%%%%%%%%%%%%%%%%%%%%%%%%%

%%%%%%%%%%%%%%%%% BODY OF PAPER %%%%%%%%%%%%%%%%%%

\section{Introduction}
Tidal interactions may have played an important role in the evolution of short-period exoplanetary systems, binary stars and planet-satellite systems. The efficiency of tidal dissipation, which is usually parameterized as the tidal quality factor $Q$ \citep{Goldreich1966Icarus}, will determine the fate of these systems. With the accumulation of observations over several decades, the observational constraint of the tidal quality factor is now possible, especially in hot Jupiter systems \citep[e.g.][]{Wilkins2017ApJ,Patra2017AJ}. However, it is still very challenging to theoretically predict the tidal quality factor owing to  intrinsic difficulties and uncertainties of the problem. %There remains several open questions yet to be explored.  

Tidal responses can be generally separated into two parts: the equilibrium tide and the dynamic tide \citep[e.g.][]{Ogilvie2014ARAA}. The equilibrium tide is a quasi-hydrostatic deformation to the gravitational pulling of the orbital companion. The dynamic tide is usually associated with different kinds of internal waves such as internal gravity waves \citep{Zahn1975AA,Savonije1983MNRAS,Goldreich1989ApJ,Goodman1998ApJ,Barker2010MNRAS,Essick2016ApJ} and inertial waves \citep{Ogilvie2004ApJ,Ogilvie2007ApJ,Wu2005ApJ,Ogilvie2009MNRA,Ogilvie2013MNRAS,Goodman2009ApJ,
Rieutord2010,Papaloizou2010MNRAS}.
The tidal dissipation associated with these hydrodynamic waves exhibits very complicated dependences on the tidal frequency \citep{Ogilvie2014ARAA}. The non-linear effect and the interactions with convection and magnetic fields may wash out some of the frequency-dependence \citep{Ogilvie2013MNRAS}, yet these effects on tides remain to be elucidated. Magnetic fields are ubiquitous in astrophysical bodies, where they can interact with tidal flows in electrically conducting fluid layers and lead to additional Ohmic dissipation. The magnetic field will modify the propagation and dissipation of hydrodynamic waves through the coupling with Alfv\'en waves. The present study aims to investigate the magnetic effects on tidally forced inertial waves. In the presence of a magnetic field and rotation, wavelike motions are hybrid inertial waves and Alfv\'en waves,  which we collectively refer to as magnetic-Coriolis waves \citep{Finlay2008}. 
%We will examine the propagation and the energy dissipation of magnetic-Coriolis waves excited by a tidal forcing in spherical shells.   

Magnetic-Coriolis (MC) waves were first investigated by \cite{Lehnert1954ApJ}. He showed the rotation can split the Alfv\'en waves into two groups: fast waves and slow waves, although this separation is vague in some parameter regimes. MC waves have been studied in different geophysical and astrophysical contexts since the pioneering study of \cite{Lehnert1954ApJ}. In geophysics, slow MC waves (also called magnetostrophic waves) are of particular interest, as these waves are thought to be important in the generation and secular variations of the Earth's magnetic field through the dynamo process in the liquid outer core \citep{Hide1966,Malkus1967JFM,Jault2008,Finlay2008,Bardsley2017}. The slow MC waves usually have frequencies much lower than the rotation frequency and are quasi-geostrophic, i.e. nearly invariant along the rotation axis. In astrophysics, studies of stellar oscillations have shown that MC waves can be excited in rotating magnetized stars \citep{Lander2010MNRAS,Abbassi2012MNRAS}. MC waves have also been observed recently in liquid metal experiments \citep{Nornberg2010, Schmitt2010GApFD,Schmitt2013EJMF}. However, tidally forced MC waves and the energy dissipation associated with MC waves have not been well studied, except for a recent study using a periodic box \citep{Wei2016ApJ}. \cite{Buffett2010} calculated the Ohmic dissipation of tidally driven (free inner-core nutation) inertial waves in the Earth's outer core, but the Lorentz force is dropped in his calculations. 

The present paper studies the propagation and the energy dissipation of tidally forced MC waves in spherical shells. We use a simplified model consisting of a rigid core and a homogeneous fluid envelope, of which the hydrodynamic responses have previously been considered \citep{Ogilvie2009MNRA,Ogilvie2013MNRAS,Rieutord2010}. Tidal responses are decomposed into non-wavelike and wavelike parts \citep{Ogilvie2013MNRAS}, and we focus on the latter in this study. The wavelike perturbations are MC waves in the presence of a magnetic field, where both the Lorentz force and the Coriolis force act as the restoring force. The linearized equations describing the wavelike perturbations are numerically solved using a pseudo-spectral method. The total energy dissipation can be contributed to by both viscous damping and Ohmic damping, but is dominated by the latter in most cases. We investigate the dependence of the total dissipation rate on the magnetic field strength, the field structure, the dissipative parameters and the tidal frequency. We also examine the frequency-averaged dissipation, which has been used to study tidal dissipation and evolution of stars recently \citep{Guenel2014AA,Mathis2015AA,Bolmont2016CeMDA,Gallet2017,Bolmont2017}. Remarkably, the frequency-averaged dissipation quantity is in good agreement with previous analytical results in the absence of magnetic fields \citep{Ogilvie2013MNRAS}. Our results suggest that the frequency-averaged tidal dissipation is insensitive to the detailed damping mechanisms, at least for the wavelike tides. This may have important implications for studying the long-term tidal evolution, as the detailed damping mechanism and dissipative properties are not well constrained in stars and planets.  

The paper is organized as follows. Section \ref{sec:maths} introduces the simplified model, the basic equations and the numerical method. Section \ref{sec:Results} presents numerical results. Section \ref{sec:Conclusion} summarizes the key findings of this study.      

%\citep{Bardsley2017}

\section{The simplified model} \label{sec:maths}
\begin{figure}
\begin{center}
\includegraphics[width=0.3 \textwidth]{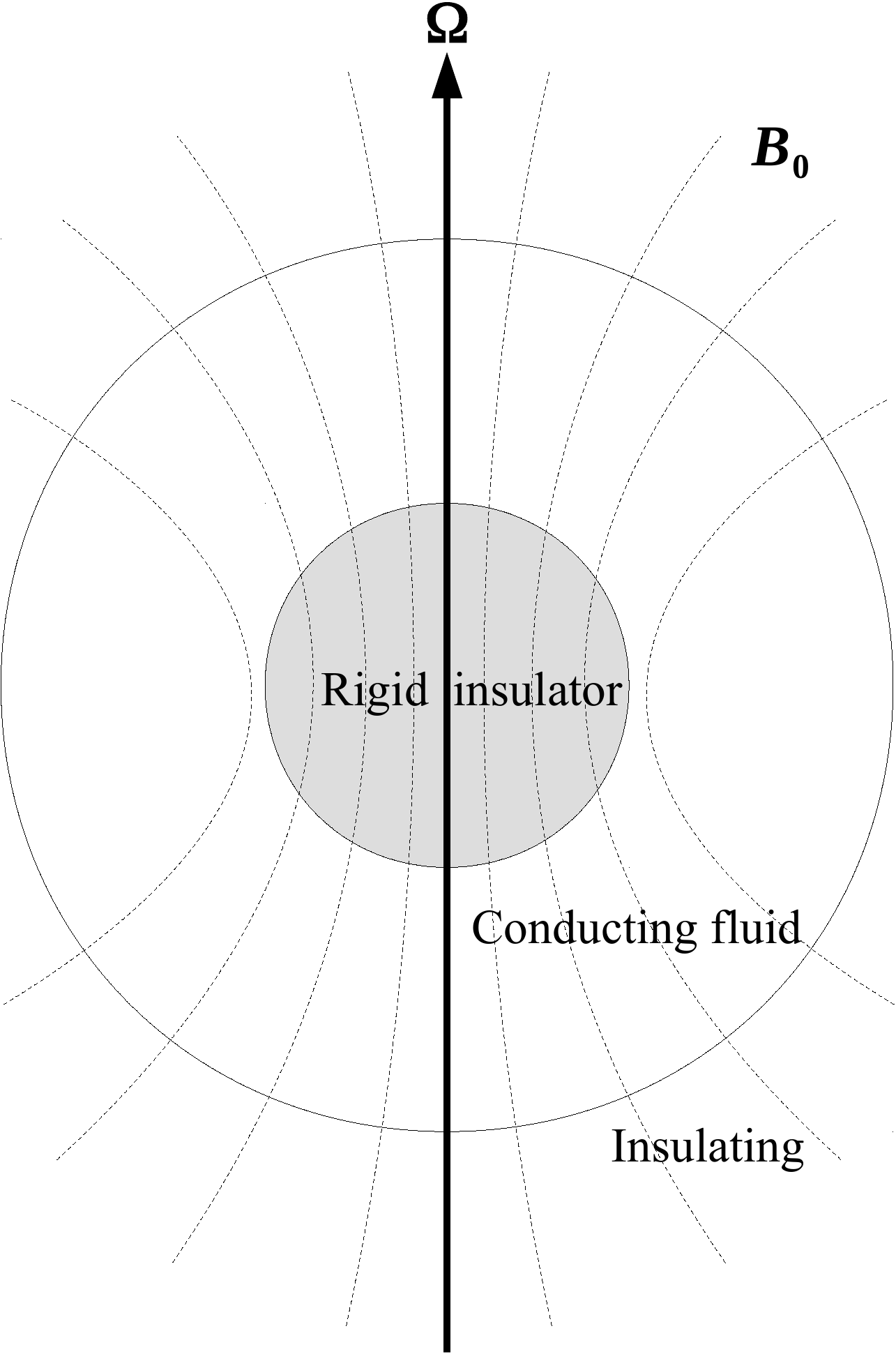}
\caption{Illustration of the model. The dashed lines are a schematic representation of the background magnetic field $\bmath{B_0}$.}
\label{fig:model}
\end{center}
\end{figure}

Our model essentially builds upon a hydrodynamic model considered by \cite{Ogilvie2009MNRA,Ogilvie2013MNRAS}. We consider a uniformly rotating spherical body consisting of a rigid inner core and a homogeneous incompressible fluid envelope. In order to take into account magnetic effects, we assume that the whole body is permeated by a steady axisymmetric magnetic field and the fluid is electrically conducting (see Fig. \ref{fig:model}). We focus on the wavelike tidal perturbations in the fluid envelope, which may be regarded as an idealized model of the convective zones of stars and giant planets as it lacks stable stratification. 
This section introduces basic equations describing the wavelike perturbations and the numerical method we used to solve these equations.      
\subsection{Basic equations}
A perfectly rigid inner core of radius $r=\alpha R$ is enclosed by a homogeneous, incompressible and electrically conducting fluid shell with a free surface at $r=R$.  We assume that the rigid core and the fluid envelope have the same density $\rho$. The whole body uniformly rotates at $\bmath \Omega=\Omega \bmath{\hat{z}}$ and is permeated by a steady magnetic field $\bmath {B_0}$. For simplicity, we assume that the background magnetic field $\bmath{B_0}$ is either a dipolar field 
\begin{equation}
\bmath{B_0}=B_0\left[\bmath{\hat{r}}\left(\frac{R}{r}\right)^3\cos \theta +\bmath{\hat{\theta}}\left(\frac{R}{r}\right)^3\frac{\sin \theta}{2} \right],
\end{equation}
or a uniform axial field
\begin{equation}
\bmath{B_0}=B_0\bmath{\hat{z}}=B_0(\bmath{\hat{r}} \cos \theta -\bmath{\hat{\theta}}\sin \theta ),
\end{equation}
where we have used spherical coordinates ($r, \theta, \phi$).
Note that for both cases, $\bmath{B_0}$ is a potential field, i.e. $\grad\btimes \bmath{B_0}=0$.

We consider the linear responses of the fluid envelope to a tidal potential $\Psi=A(r/R)^l Y_l^m(\theta,\phi)\mathrm{e}^{-\mathrm{i}\omega t} $, where $Y_l^m(\theta,\phi)$ is a spherical harmonic and $\omega$ is the tidal frequency in the rotating frame. The tidal responses can be decomposed into non-wavelike and wavelike parts as introduced by \cite{Ogilvie2013MNRAS}. In the absence of a magnetic field, the non-wavelike part represents the instantaneous response to the tidal potential, while the wavelike part is in the form of inertial waves excited by the effective force \citep{Ogilvie2013MNRAS}
\begin{equation}\label{eq:forcing}
\bmath f = 2 \bmath{\Omega} \btimes \grad[X_l(r)Y_l^m(\theta,\phi)]\mathrm{e}^{-\mathrm{i}\omega t},
\end{equation}
where $X_l$ is associated with the non-wavelike motions. 
This effective force is not generally irrotational and results from the failure of the non-wavelike tide to satisfy the equation of motion when the Coriolis force is included. 
For a homogeneous incompressible fluid of the same density as that of the rigid core, $X_l(r)$ is given as \citep{Ogilvie2013MNRAS}
\begin{equation}
X_l(r)=C_l\left[ \left(\frac{r}{R}\right)^l+\alpha^{2l+1}\frac{l}{l+1}\left(\frac{R}{r}\right)^{l+1}\right].
\end{equation}
The constant $C_l$ is given by
\begin{equation}
C_l=\frac{\mathrm{i}\omega(2l+1)R^3A}{2l(l-1)(1-\alpha^{2l+1})GM},
\end{equation} 
where $G$ is the gravitational constant and $M$ is the total mass of the body.
In this paper, we consider only the dominant tidal component of $l=2$ and $m=2$, unless otherwise specified. 

In the presence of a magnetic field, we assume that the large-scale non-wave-like part is unchanged, and that the wave-like perturbations are still driven by the effective force given in equation (\ref{eq:forcing}), but we allow the magnetic field to affect the small-scale wave-like motions through the induction equation and the Lorentz force.  
 The linearized equations governing the wavelike velocity perturbation $\bmath u$ and magnetic field perturbation $\bmath b$  in the rotating frame can be written as 
\begin{equation} \label{eq:NS_dimen}
 {-\mathrm{i}\omega }\bmath{u}+2\bmath{\Omega}\btimes \bmath u=-\frac{1}{\rho}\grad p +\frac{1}{\rho \mu_0}(\grad\btimes \bmath b)\btimes \bmath{B_0}+\nu \grad^2 \bmath u+ \bmath f,
\end{equation}
\begin{equation} \label{eq:B_dimen}
{-\mathrm{i}\omega }\bmath{b}=\grad\btimes( \bmath u \btimes \bmath{B_0})+\eta \grad^2 \bmath b,
\end{equation}
\begin{equation}\label{eq:divu0}
\div \bmath u=0,
\end{equation}
\begin{equation}\label{eq:divb0}
\div \bmath b=0,
\end{equation}
where $\nu$ is the fluid viscosity, $\mu_0$ is the magnetic permeability and $\eta$ is the magnetic diffusivity.  Equation (\ref{eq:NS_dimen}) is the Navier-Stokes equation including the Coriolis force and the Lorentz force. Equation~(\ref{eq:B_dimen}) is the magnetic induction equation.

Using $R$, $\Omega^{-1}$, $B_0$ as units of length, time and magnetic field strength, we get the non-dimensional equations:
\begin{equation}\label{eq:Nod_Ns}
{-\mathrm{i}\omega }\bmath{u}+2\bmath{\hat{z}}\btimes \bmath u=-\grad p+Le^2 (\grad\btimes \bmath b)\btimes \bmath{B_0}+E_k \grad^2 \bmath u +\bmath f,
\end{equation}
\begin{equation}\label{eq:Nod_Indu}
{-\mathrm{i}\omega }\bmath{b}=\grad\btimes( \bmath u \btimes \bmath{B_0})+E_m \grad^2 \bmath b,
\end{equation}
where (and whereafter) $\omega$, $\bmath u$, $p$, $\bmath b$, $\bmath B_0$ and $\bmath f$ denote the corresponding non-dimensional quantities. 
The dimensionless parameters in equations~(\ref{eq:Nod_Ns}-\ref{eq:Nod_Indu}) are the Lehnert number $Le$, the Ekman number $E_k$ and the magnetic Ekman number $E_m$:
\begin{equation}
Le=\frac{B_0}{\sqrt{\rho \mu_0}\Omega R}, \quad E_k=\frac{\nu}{\Omega R^2}, \quad E_m=\frac{\eta}{\Omega R^2}. 
\end{equation}
The Lehnert number $Le$ measures the strength of the background magnetic field with respect to rotation, i.e. the ratio  between the Alf\'ven velocity $B_0/\sqrt{\rho \mu_0}$ and the rotation velocity $\Omega R$ at the equator. The Ekman number $E_k$ is the ratio between the rotation time scale $\Omega^{-1}$ and the viscous time scale $R^{2}/\nu$, while the magnetic Ekman number $E_m$ is the ratio between the rotation time scale $\Omega^{-1}$ and the magnetic diffusion time scale $R^{2}/\eta$.  
The last two parameters are related by the magnetic Prandtl number
\begin{equation}
Pm=\frac{E_k}{E_m}=\frac{\nu}{\eta}.
\end{equation}

We set the dimensionless forcing term as
\begin{equation}\label{eq:forcing_nd}
\bmath f= \bmath{\hat{z}}\btimes \grad[(r^2+\alpha^5/r^3)Y_2^2(\theta,\phi)](1-\alpha^5)^{-1}\mathrm{e}^{-\mathrm{i}\omega t}, 
\end{equation}
for $l=2$ and $m=2$.
%Alternatively, the effect of the magnetic field can be also measured by the Elsasser number:
%\begin{equation}
%\Lambda=\frac{B_0^2}{\Omega \rho \mu_0 \eta}=\frac{Le^2Pm}{E_k}=\frac{Le^2}{E_m}.
%\end{equation}

In order to minimize the viscous and electromagnetic couplings between the fluid layer and the rigid core, we use the stress-free boundary condition for the velocity $\bmath u$ and the insulating boundary condition for the magnetic field $\bmath b$ at both inner and outer boundaries. 
The stress-free boundary condition implies that the tangential component of the viscous stress should vanish: 
\begin{equation} \label{eq:stress_BC}
u_r=\frac{\partial}{\partial r}\left(\frac{u_\theta}{r}\right)=\frac{\partial}{\partial r}\left(\frac{u_\phi}{r}\right)=0
\end{equation}
The insulating boundary condition indicates that no electric current can go through the boundaries:
 \begin{equation} \label{eq:current_BC}
 (\grad \btimes \bmath b)_r=0,
 \end{equation}
and the magnetic field in the conducting fluid should also match a potential field $\bmath{B}^{e}=-\grad P$ in the exterior insulating regions, where $P$ is a scalar potential. This condition can be readily expressed in terms of spherical harmonics (see Appendix \ref{App:Projected}).

The viscous dissipation rate in dimensionless form is
\begin{equation}\label{eq:Dvis}
D_{\mathrm{vis}}=\frac{1}{2}E_k \int_V \mathrm{Re}[(\grad^2 \bmath u)\bcdot \bmath u^*] \mathrm{d}V,
\end{equation}
and the dimensionless Ohmic dissipation rate is
\begin{equation}\label{eq:Dohm}
D_{\mathrm{ohm}}=\frac{1}{2}Le^2E_m \int_V |\grad \btimes \bmath b|^2 \mathrm{d}V,
\end{equation} 
where integrals are evaluated over the fluid domain. It can be shown from the integrated energy equation that the total energy dissipation rate equals the power input by the tidal forcing $\bmath f$
\begin{equation}\label{eq:Dtot}
D_{\mathrm{vis}}+D_{\mathrm{ohm}}\equiv\frac{1}{2}\int_V \mathrm{Re}[\bmath f \bcdot \bmath u^*] \mathrm{d}V,
\end{equation}
where $\bmath u^*$ is the complex conjugate of $\bmath u$. 
\subsection{Numerical method}{\label{subsec:numerical}}
Equations (\ref{eq:Nod_Ns}-\ref{eq:Nod_Indu}) are solved using a pseudo-spectral method. We use a spheroidal-toroidal decomposition and then project the equations on to spherical harmonics in a similar way as in \cite{Rincon2003AA}. The velocity and magnetic field perturbations are expanded as: 
\begin{equation} \label{eq:u_RST}
\bmath u=	\sum u_l^m(r) \bmath R_l^m+\sum v_l^m(r) \bmath S_l^m+ \sum w_l^m(r) \bmath T_l^m ,
\end{equation}
\begin{equation} \label{eq:b_RST}
\bmath b=	\sum a_l^m(r) \bmath {R}_l^m+\sum b_l^m(r) \bmath S_l^m+ \sum c_l^m(r) \bmath T_l^m ,
\end{equation}
with the summation is carried out over integers $l\geq m\geq 0$. Here $\bmath R_l^m$,$\bmath S_l^m$, $\bmath T_l^m$ are vector spherical harmonics:
\begin{equation}
\bmath R_l^m=Y_l^m(\theta,\phi) \bmath{\hat{r}}, \ \bmath S_l^m=r \grad Y_l^m(\theta,\phi) , \  \bmath T_l^m=r \grad \btimes  \bmath R_l^m.
\end{equation}
For the divergence-free fields, the first two terms in equations (\ref{eq:u_RST}-\ref{eq:b_RST}) are also referred to as the poloidal part and the last term is the toroidal part. 
The divergence-free conditions of $\bmath u$ and $\bmath b$ (equations \ref{eq:divu0}-\ref{eq:divb0}) are satisfied by
\begin{equation}
v_l^m=\frac{1}{l(l+1)r}\frac{\mathrm{d}(r^2 u_l^m)}{\mathrm{d} r},
\end{equation}
\begin{equation}\label{eq:divblm0}
b_l^m=\frac{1}{l(l+1)r}\frac{\mathrm{d}(r^2 a_l^m)}{\mathrm{d} r}.
\end{equation}
The equations projected on to spherical harmonics are given in Appendix \ref{App:Projected}. We can see that equations (\ref{eq:A1}-\ref{eq:A4}) are decoupled for each $m$ owing to the axisymmetric rotation and magnetic field $\bmath{B_0}$. The Coriolis force and the magnetic field only couple the neighbouring spherical harmonics $l-1$ and $l+1$. Numerically, the system is truncated at a spherical harmonical degree $L$. In radial direction, we use Chebyshev collocation on $N+1$ Gauss-Lobatto nodes. We use a typical truncation of $L=N=400$ in most of calculations, 
but higher resolutions up to $L=N=600$ are also used for a few more demanding calculations (when $E_m\leq 10^{-5}$ and $Le\leq 10^{-4}$).  
The numerical discretization leads to linear equations involving a large block-tridiagonal matrix, which is solved using the standard direct method based on LU factorization. 

The stress-free boundary condition at the inner and outer boundaries becomes
\begin{equation}
u_l^m=\frac{\mathrm{d}}{\mathrm{d}r}\left(\frac{v_l^m}{r}\right)=\frac{\mathrm{d}}{\mathrm{d}r}\left(\frac{w_l^m}{r}\right)=0.
\end{equation}
The insulating boundary condition requires vanishing toroidal field, i.e. $c_l^m=0$, at the inner and outer boundaries. The poloidal field needs to match a potential field, leading to (see Appendix \ref{App:Projected})
 \begin{equation}
\frac{d a_l^m}{dr}-\frac{l-1}{r}a_l^m=0,
\end{equation}
at the inner boundary and 
\begin{equation}
\frac{d a_l^m}{dr}+\frac{l+2}{r}a_l^m=0,
\end{equation}
at the outer boundary. 
%The top and bottom rows of each blocks are replaced by the corresponding boundary conditions.

The dissipation rates in equations (\ref{eq:Dvis}-\ref{eq:Dohm}) can be reduced to integrals in radius only using the orthogonality of spherical harmonics:
%\begin{multline}
%D_{\mathrm{vis}}=\frac{1}{2}E_k \int_{\alpha}^1 \sum_{l=m}^{L}l^2(l+1)^2|w_l^m(r)|^2 
%+l(l+1)\left|\frac{\mathrm{d}[rw_l^m(r)]}{\mathrm{d}r}\right|^2 \\+l(l+1)\left|u_{l}^m(r)- \frac{\mathrm{d}[rv_l^m(r)]}{\mathrm{d}r}\right|^2 \mathrm{d} r,
%\end{multline}
\begin{multline}
D_{\mathrm{vis}}=\frac{1}{2}E_k \int_{\alpha}^1 \sum_{l=m}^{L} l(l+1)\left|u_{l}^m(r)+r^2\frac{\mathrm{d}[v_l^m(r)/r]}{\mathrm{d}r}\right|^2 \\
+l(l+1)\left|r^2\frac{\mathrm{d}[w_l^m(r)/r]}{\mathrm{d}r}\right|^2+3\left| r \frac{\mathrm{d} u_l^m(r)}{\mathrm{d}r}  \right|^2  \\
+(l-1)l(l+1)(l+2)\left(|v_l^m(r)|^2 +|w_l^m(r)|^2\right) 
 \mathrm{d} r,
\end{multline}
\begin{multline}
D_{\mathrm{ohm}}=\frac{1}{2}Le^2E_m \int_{\alpha}^1 \sum_{l=m}^{L}l^2(l+1)^2|c_l^m(r)|^2  \\
+l(l+1)\left|\frac{\mathrm{d}[rc_l^m(r)]}{\mathrm{d}r}\right|^2+l(l+1)\left|a_{l}^m(r)- \frac{\mathrm{d}[rb_l^m(r)]}{\mathrm{d}r}\right|^2 \mathrm{d} r.
\end{multline}
The integrals are evaluated by a Chebyshev quadrature formula which uses the function values at the collocation points. 
We also calculate the total dissipation rate using equation (\ref{eq:Dtot}), which involves only spectral coefficients  of  $l=1$ and $l=3$. This is because the effective forcing term projected onto spherical harmonics has only $l=1$ and $l=3$ components for the $l=2$ tidal forcing (see equations~(\ref{eq:fl}-\ref{eq:fl2}) in Appendix \ref{App:Projected}), and because of the orthogonality of spherical harmonics. The identity (\ref{eq:Dtot}) can be used to check the numerical accuracy and convergence. Our numerical code is also validated by comparing with some of the results in \cite{Rincon2003AA} and \cite{Ogilvie2009MNRA}.
\subsection{Dispersion relation of magnetic-Coriolis waves}
Before presenting our numerical results, let us briefly recall the dispersion relation of magnetic-Coriolis waves, which is useful for the discussion of some results. Substituting the plane wave ansatz $\bmath u, \, \bmath b \propto \mathrm{e}^{\mathrm{i}(\bmath k \bcdot \bmath r-\omega t)}$ into equations (\ref{eq:NS_dimen}-\ref{eq:B_dimen}), and neglecting the diffusive terms and the forcing term, we can obtain the dispersion relation of magnetic-Coriolis waves in a uniform field $\bmath B_0$ \citep[e.g.][]{Finlay2008}:
\begin{equation} \label{eq:dispersion}
\omega= \pm \frac{\bmath \Omega \bcdot \bmath k}{|\bmath k|}\pm\left(\frac{(\bmath \Omega \bcdot \bmath k)^2}{|\bmath k|^2}+\frac{(\bmath B_0 \bcdot \bmath k)^2}{\rho \mu_0}\right)^{1/2}.
\end{equation}
In the absence of a magnetic field, i.e. $\bmath B_0=0$, equation (\ref{eq:dispersion}) recovers the dispersion relation of inertial waves
\begin{equation}
\omega=\pm2 \frac{\bmath \Omega \bcdot \bmath k}{|\bmath k|},
\end{equation}
which exist only when $|\omega|<2 \Omega$. The group velocity of inertial waves is 
\begin{equation}
\bmath{V_g}=\pm2 \frac{\bmath k \btimes (\bmath \Omega \btimes \bmath k)}{|\bmath k|^3}.
\end{equation} 
In the absence of rotation, i.e. $\bmath \Omega=0$, the dispersion relation of Alfv\'en waves is obtained:
\begin{equation}
\omega=\bmath{V_a}\bcdot \bmath k, 
\end{equation}
where $\bmath{V_a}=\bmath{B_0}/\sqrt{\rho \mu_0}$ is the group velocity. 

The propagation of MC waves is more complicated, depending on the Lehnert number which measures the importance of the magnetic field with respect to the rotation. The dispersion relation (\ref{eq:dispersion}) in dimensionless form can be written as
\begin{equation} \label{eq:dispersion_nd}
\omega= \pm \frac{\bmath {\hat z} \bcdot \bmath k}{|\bmath k|}\pm\left(\frac{(\bmath {\hat{z}} \bcdot \bmath k)^2}{|\bmath k|^2}+Le^2 k_B^2\right)^{1/2},
\end{equation}
where $k_B$ is the wavenumber along the magnetic field $\bmath{B_0}$.
\section{results} \label{sec:Results}
\subsection{Overview}
\begin{figure}
\begin{center}
(a)\includegraphics[width=0.48 \textwidth,trim={0cm 1.55cm 0cm 1.5cm},clip]{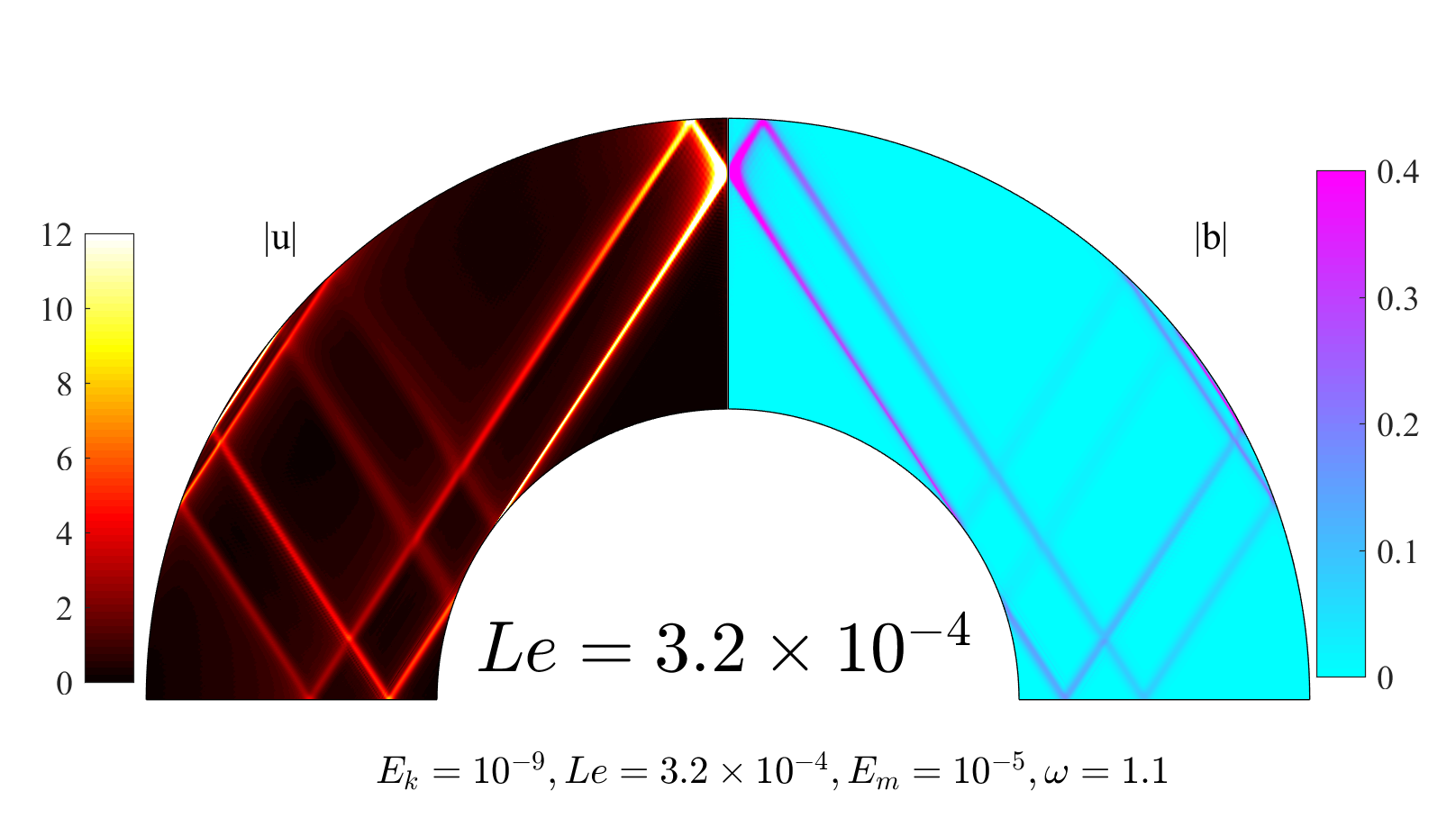} \\ 
(b)\includegraphics[width=0.48 \textwidth,trim={0cm 1.55cm 0cm 1.5cm},clip]{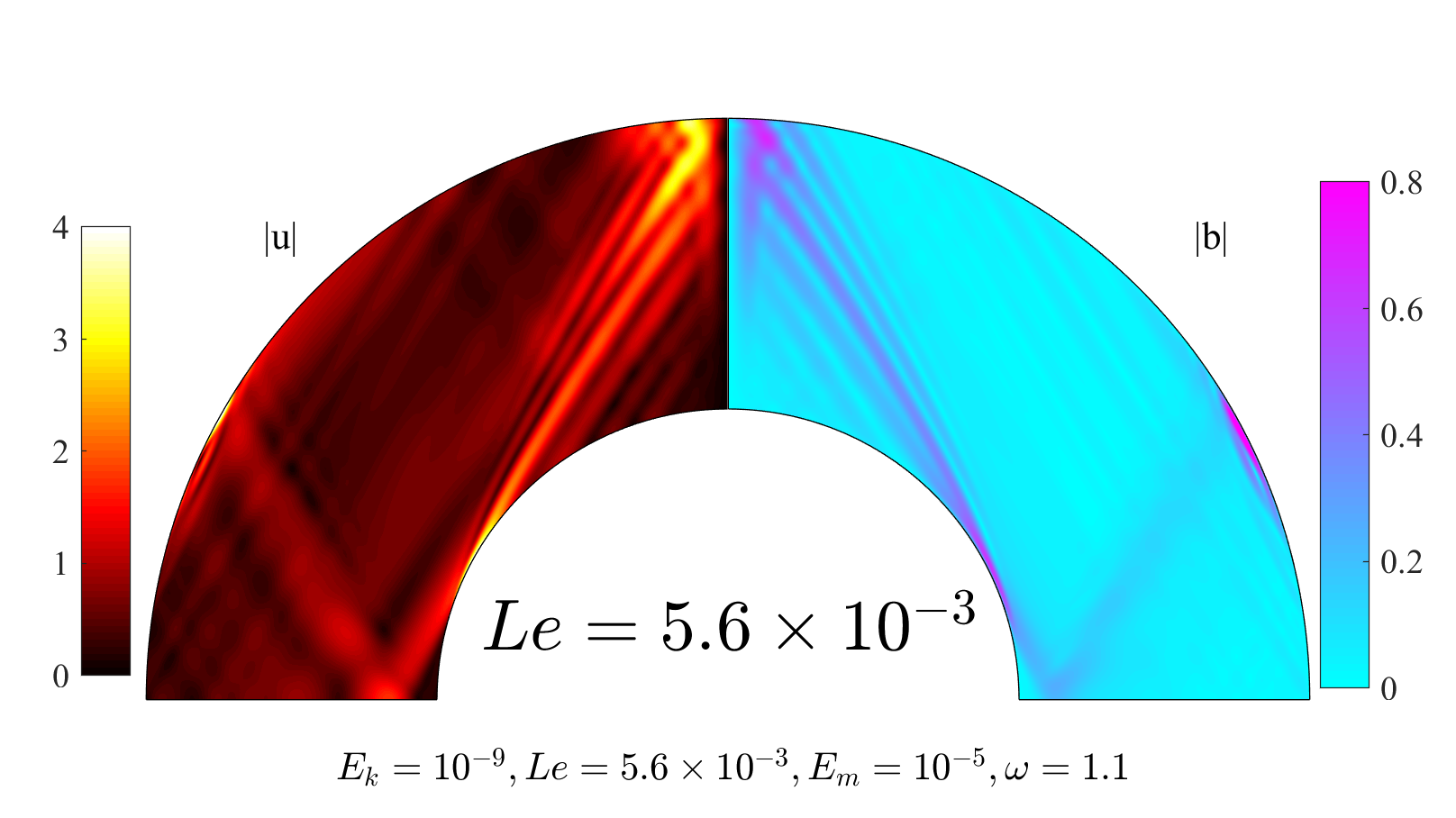}\\
(c)\includegraphics[width=0.48 \textwidth,trim={0cm 1.55cm 0cm 1.5cm},clip]{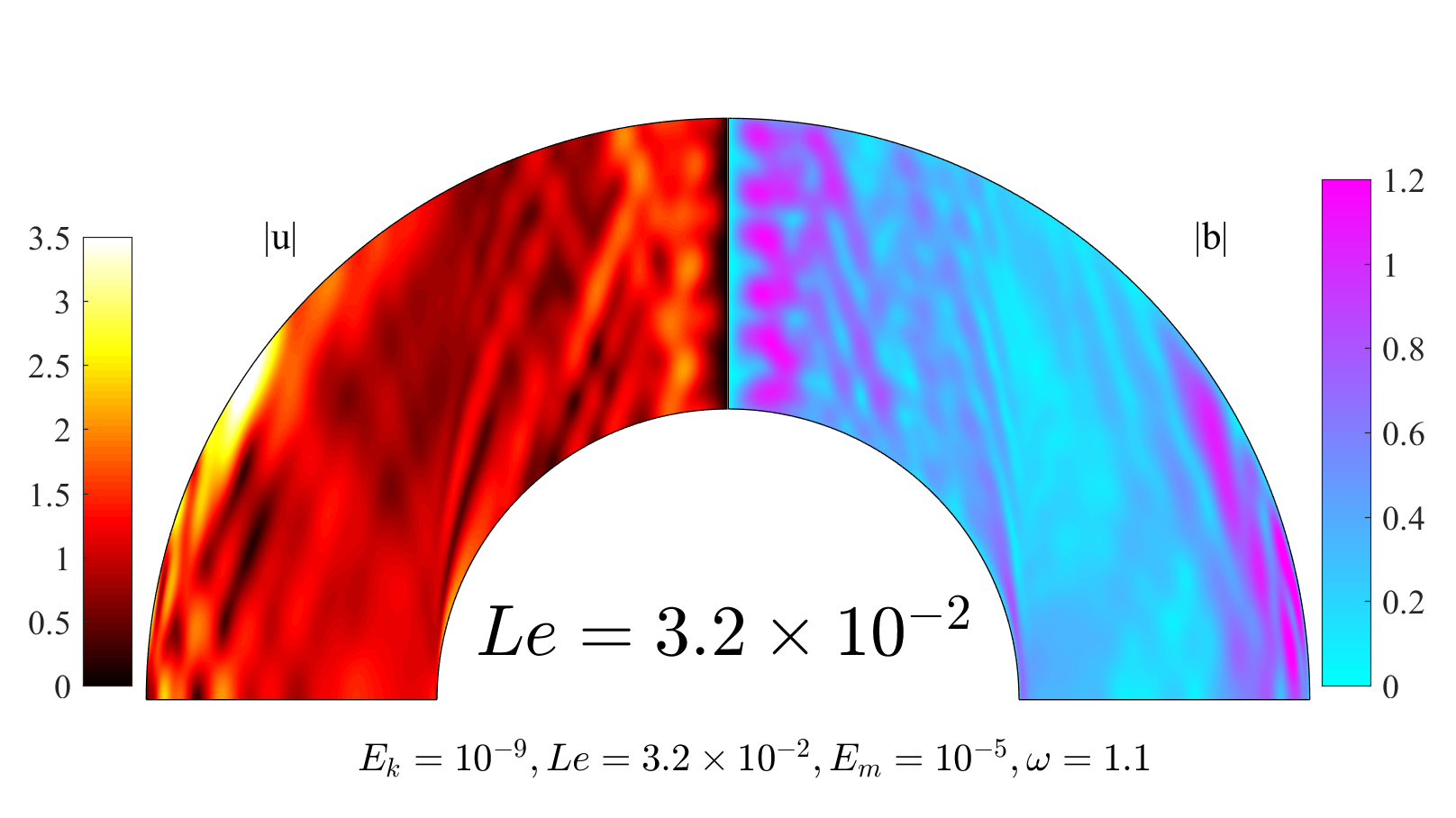}\\
(d)\includegraphics[width=0.48 \textwidth,trim={0cm 1.55cm 0cm 1.5cm},clip]{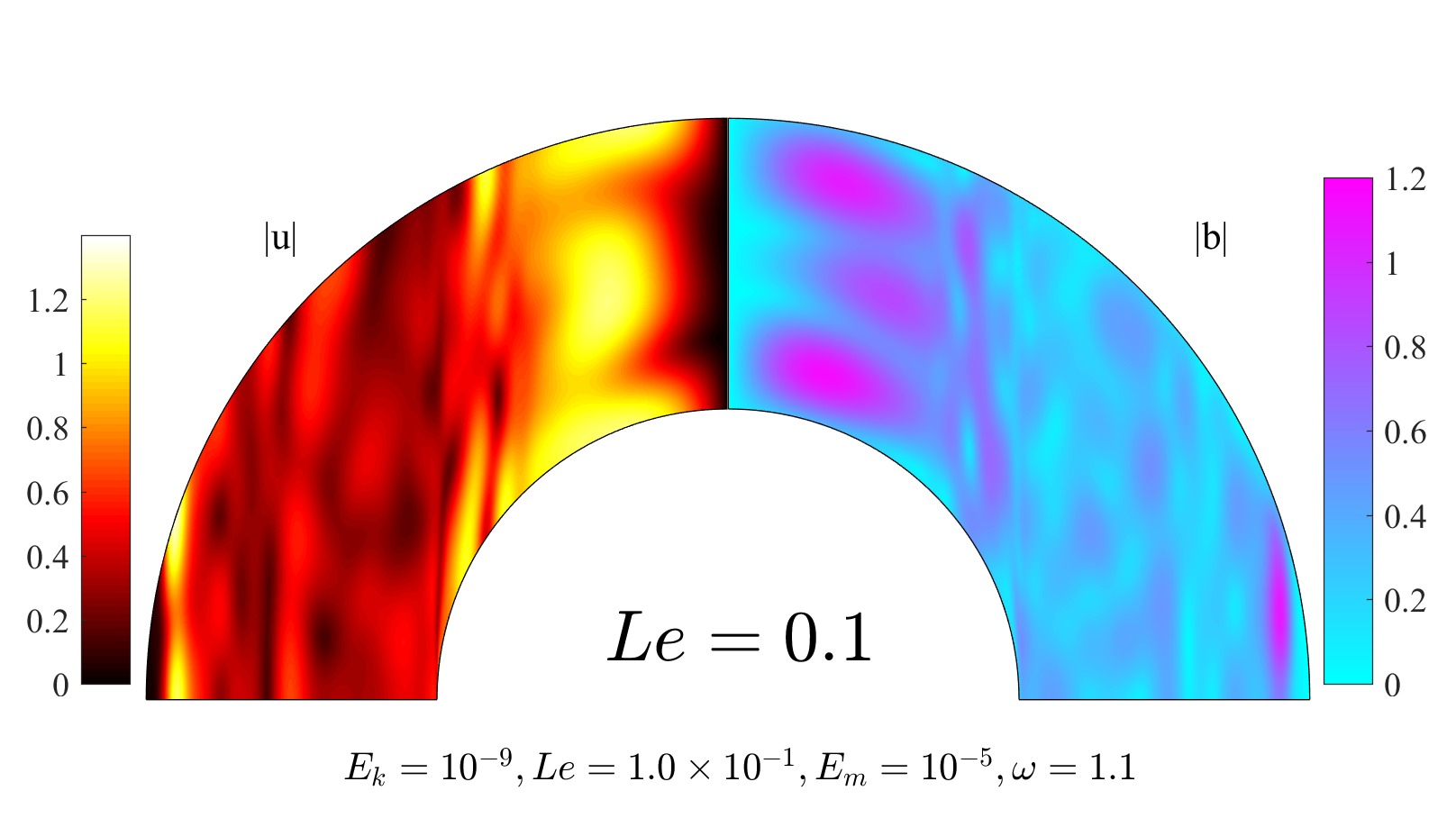}\\
(e)\includegraphics[width=0.48 \textwidth,trim={0cm 1.55cm 0cm 1.5cm},clip]{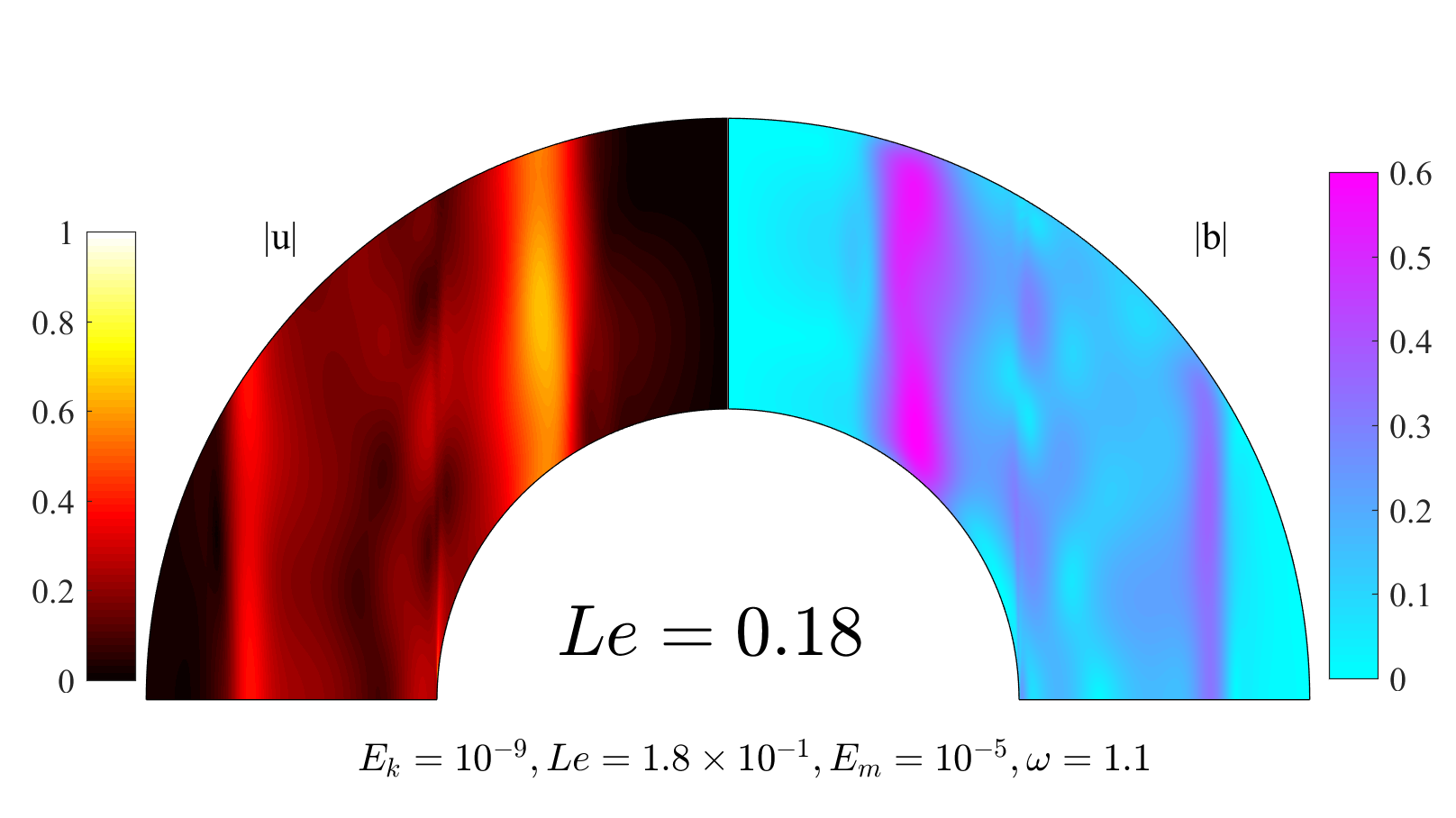} \\
(f)\includegraphics[width=0.48 \textwidth,trim={0cm 1.55cm 0cm 1.5cm},clip]{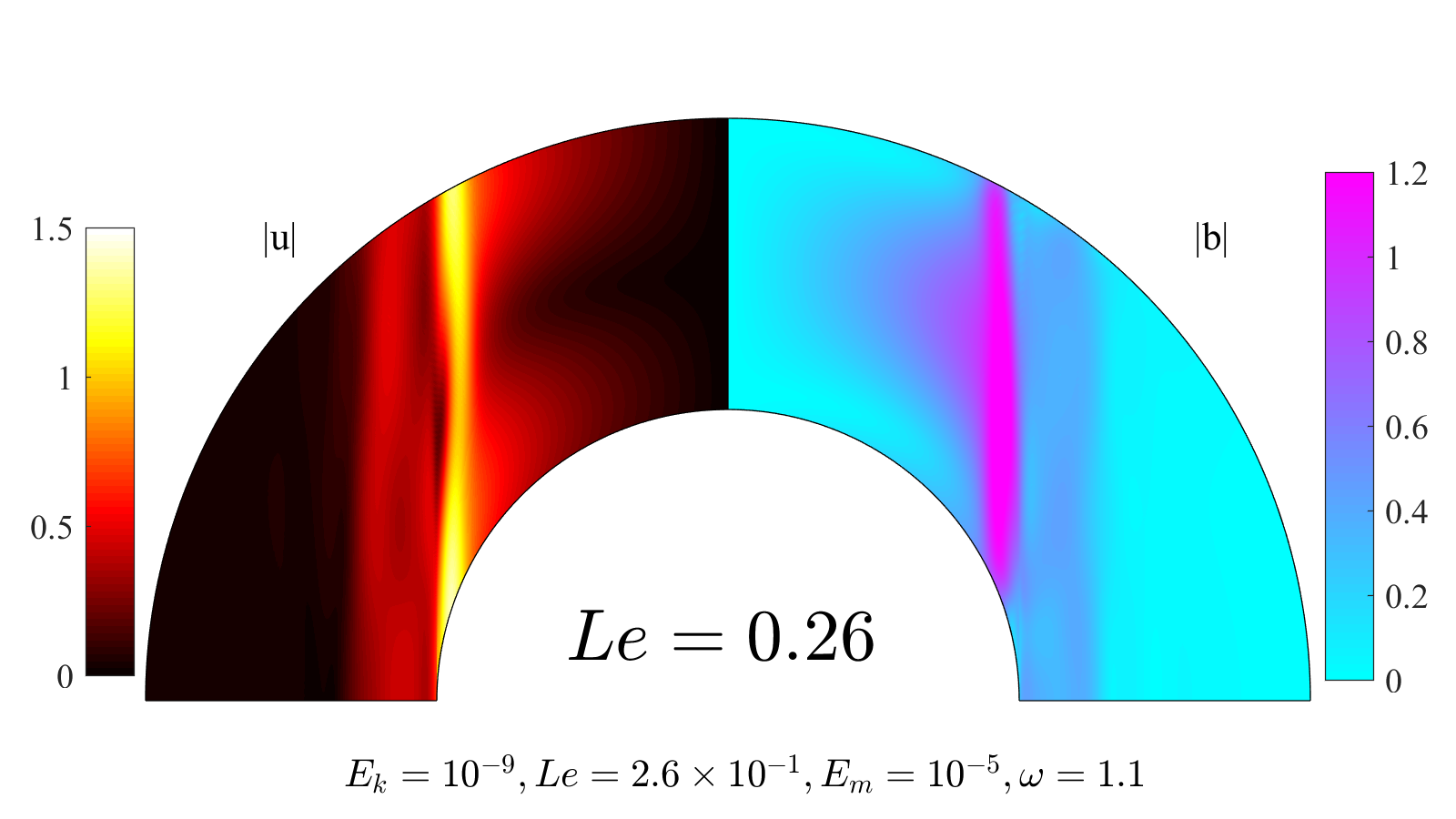} 
 \caption{Structure of the velocity perturbation $|\bmath u|$ and the magnetic field perturbation $|\bmath b|$ in the meridional plane with an axial field $\bmath{B_0}$ at different $Le$.  $E_k=1.0\times 10^{-9}$, $E_m=1.0\times 10^{-5}$, $\omega=1.1$, $\alpha=0.5$. Note that the colour scales may be different for different $Le$.}
 \label{fig:Bz} 
\end{center}
\end{figure}
In this section, we show a general overview of the spatial structure and the dissipation rate of tidally forced MC waves by varying the Lehnert number $Le$. We consider a case of the radius ratio $\alpha=0.5$ and the tidal frequency $\omega=1.1$, of which the hydrodynamic response has been studied in detail \citep{Ogilvie2009MNRA}. In the absence of a magnetic field, inertial waves propagate along the characteristics (at a fixed angle with respect to the rotation axis) and form two simple wave attractors after multiple reflections (see Fig. 9 in \cite{Ogilvie2009MNRA}). The dissipation rate associated with inertial wave attractors is independent of the viscosity (the Ekman number) provided that the Ekman number is asymptotically small \citep{Ogilvie2005JFM,Ogilvie2009MNRA}. We use this case as a reference, mainly because of its relatively simple hydrodynamic responses, to examine the effects of magnetic fields.  

Fig. \ref{fig:Bz} shows the velocity perturbation $|\bmath u|$ and the magnetic field perturbation $|\bmath b|$ in the meridional plane in the presence of an axial magnetic field with various values of the Lehnert number $Le$.      The dissipative parameters are fixed at $E_k=10^{-9}$ and $E_m=10^{-5}$, meaning that the magnetic Prandtl number $Pm=10^{-4}$. When the Lehnert number is sufficiently small ($Le\le O(E_m^{2/3})$ as we shall show later), MC waves retain the rays of inertial waves, leading to the wave attractors as for purely inertial waves (Fig \ref{fig:Bz} a). Weak magnetic field perturbations are induced along the attractors by the velocity perturbations, but the Lorentz force has negligible influence on the propagation of waves.
As the Lehnert number is gradually increased, the perturbations do not concentrate on the wave attractors any more, because the Lorentz force starts to play a part. In Fig \ref{fig:Bz} (b), however, we can still see the predominant effect of the rotation as the perturbations are mainly organized along the characteristics of inertial waves, but slightly modified. 

As we increase the Lehnert number further, the effect of rotation becomes less visible and the perturbations spread out to the whole fluid domain (Fig \ref{fig:Bz} c). At certain values of the Lehnert number, e.g. $Le=0.1$ for this case, we observe some large-scale structures in the polar region in Fig \ref{fig:Bz} (d). These structures may be associated with eigen-modes of the system. We shall show more examples of such structures at different frequencies in section \ref{subsec:fre_scan}. 
%It is well-known that the eigen value problem of inertial waves  in a spherical shell is ill-posed, so there is no general smooth inertial modes exists except a few special modes, e.g. purely toroidal modes \citep{Rieutord2001}. We suspect that there might exist some smooth magnetic-Coriolis modes at some given frequencies and Lehnert numbers. However, the formal mathematical demonstration is beyond the scope of this paper. We shall show a few more examples of such modes at different frequencies in section \ref{subsec:fre_scan}.  

At relatively large values of $Le$, i.e. $Le>0.1$, the magnetic effect become predominant as we can see from Fig. \ref{fig:Bz}(e-f) that the perturbations concentrate along certain magnetic field lines. In this regime, the perturbations are essentially in the form of Alfv\'en waves, where the magnetic tension acts as the restoring force. Each magnetic field line can be analogous to a string, which has a natural frequency depending on the length and strength of the field line. Perturbations mainly concentrate along certain field lines, where a resonance may occur if the tidal frequency matches the natural frequency of the field line.

\begin{figure}
\begin{center}
(a)\includegraphics[width=0.48 \textwidth,trim={0cm 1.55cm 0cm 1.5cm},clip]{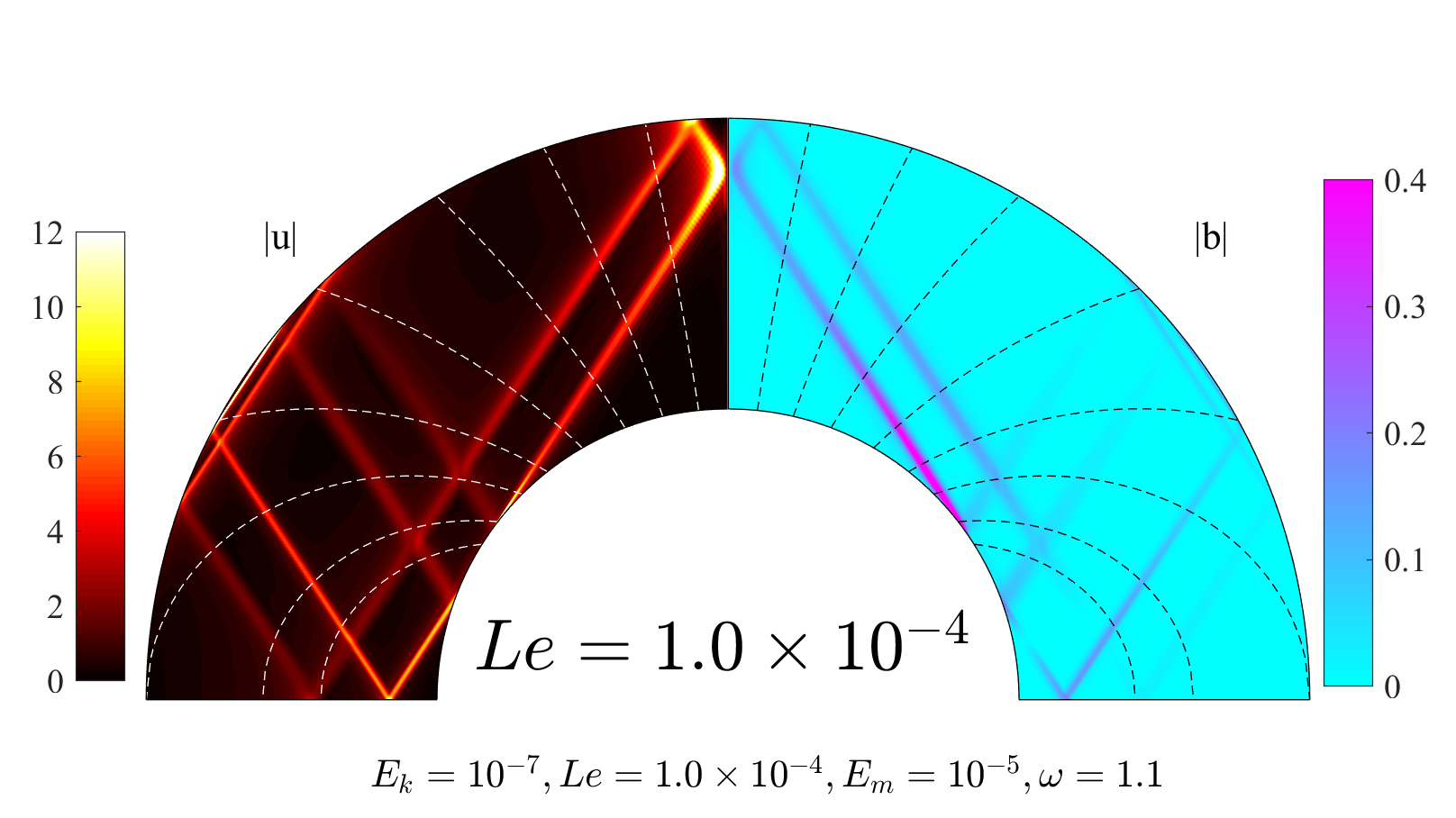}  \\
(b)\includegraphics[width=0.48 \textwidth,trim={0cm 1.55cm 0cm 1.5cm},clip]{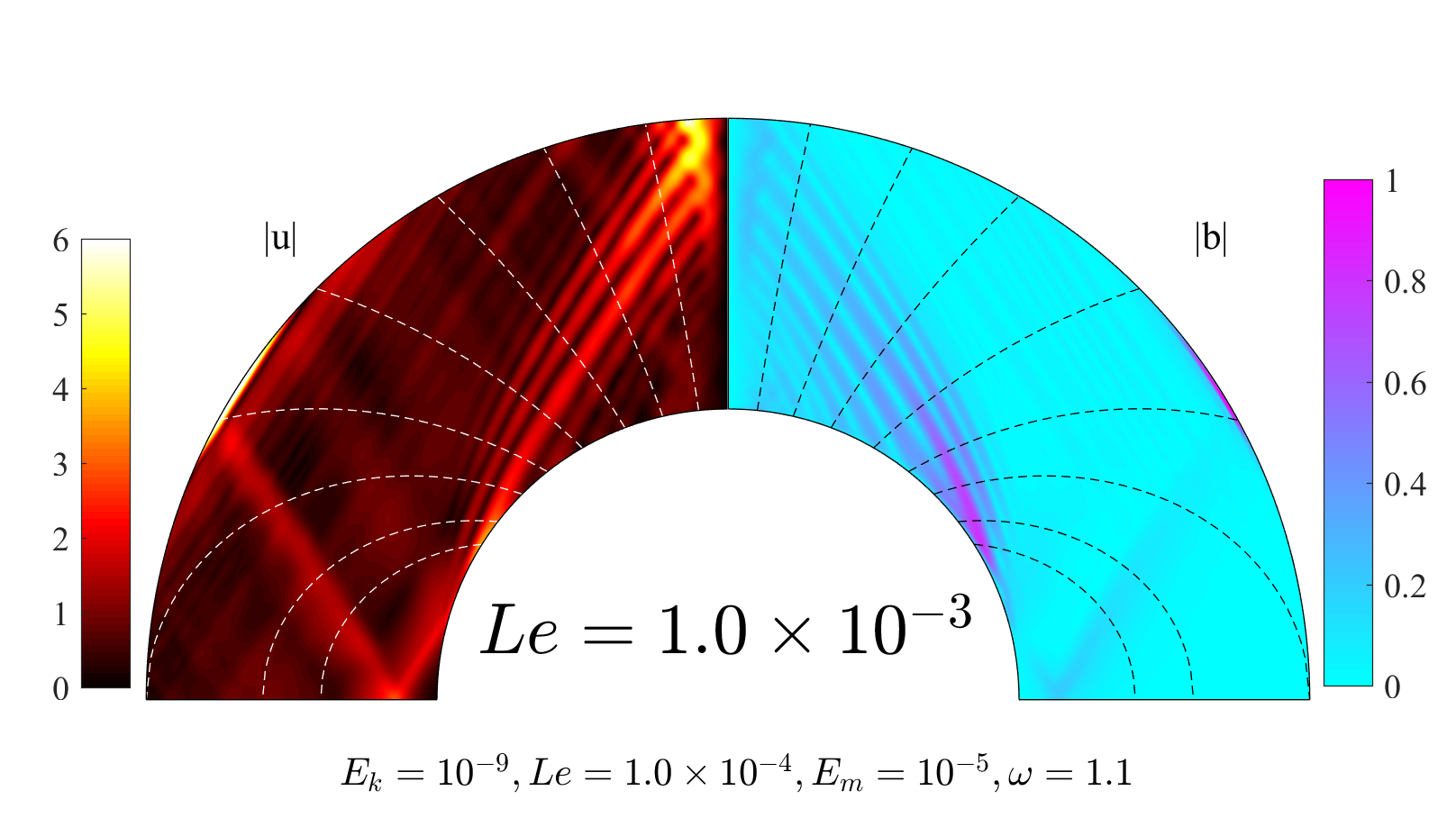}\\
(c)\includegraphics[width=0.48 \textwidth,trim={0cm 1.55cm 0cm 1.5cm},clip]{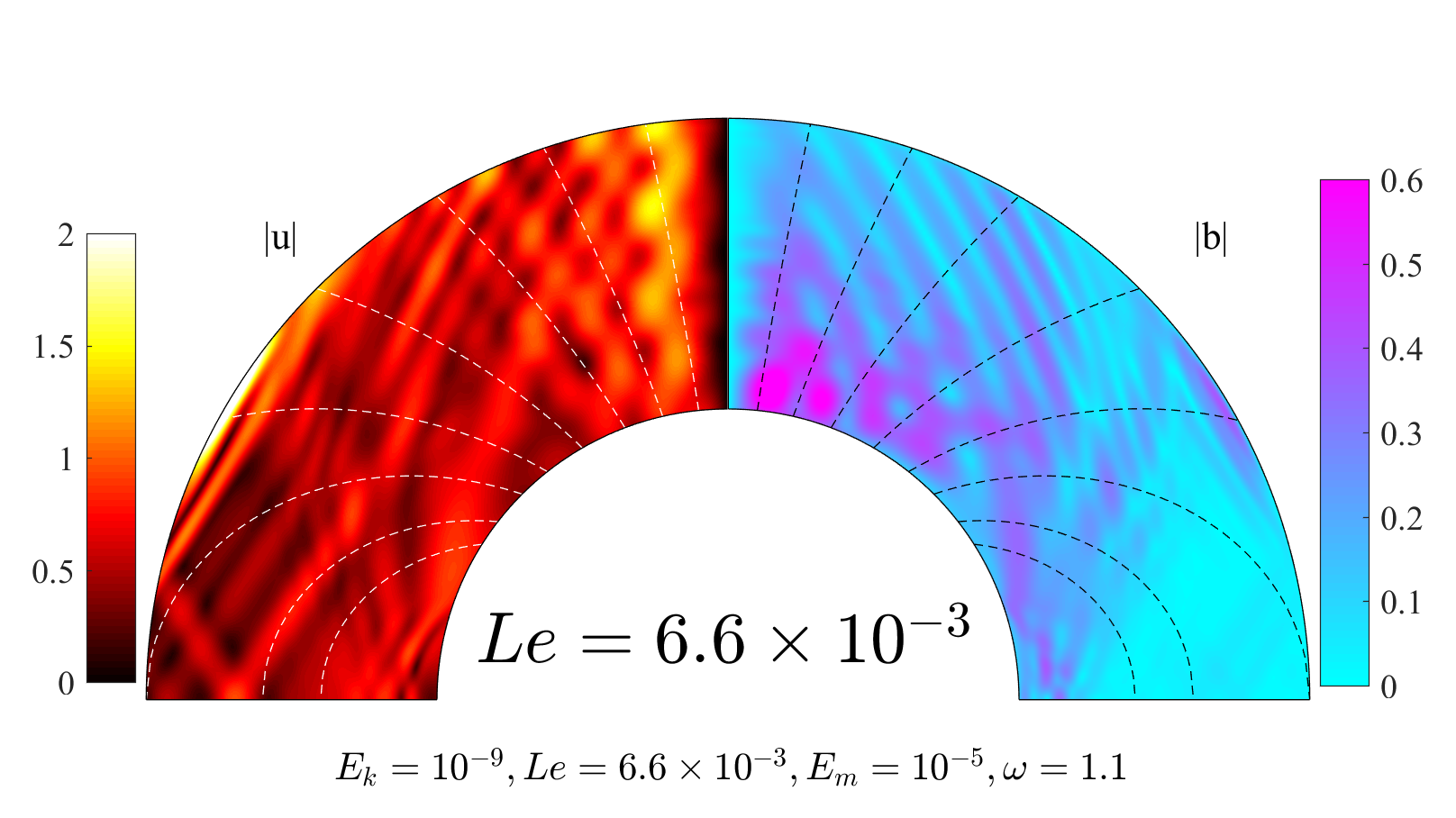}\\
(d)\includegraphics[width=0.48 \textwidth,trim={0cm 1.55cm 0cm 1.5cm},clip]{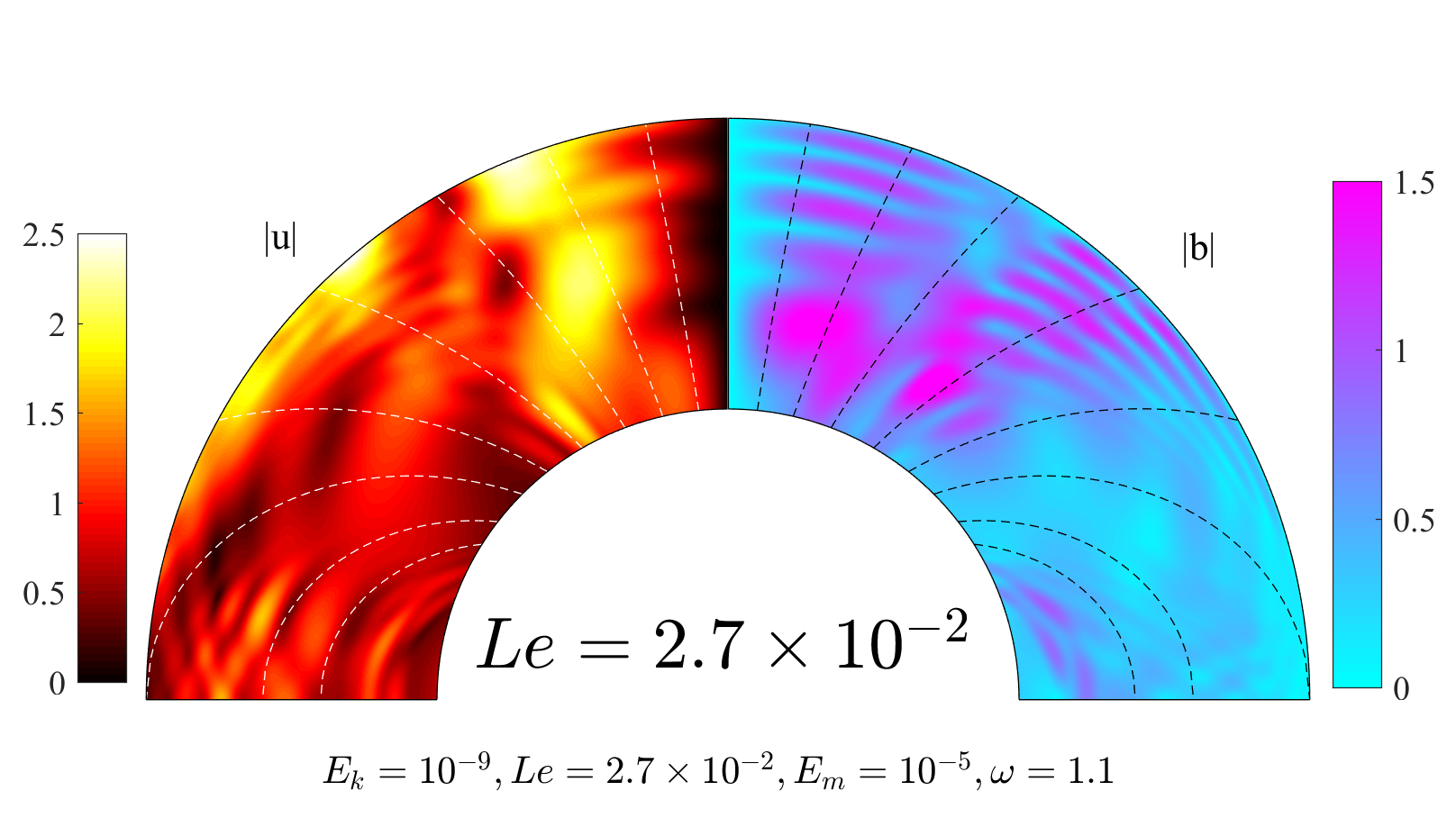}\\
(e)\includegraphics[width=0.48 \textwidth,trim={0cm 1.55cm 0cm 1.5cm},clip]{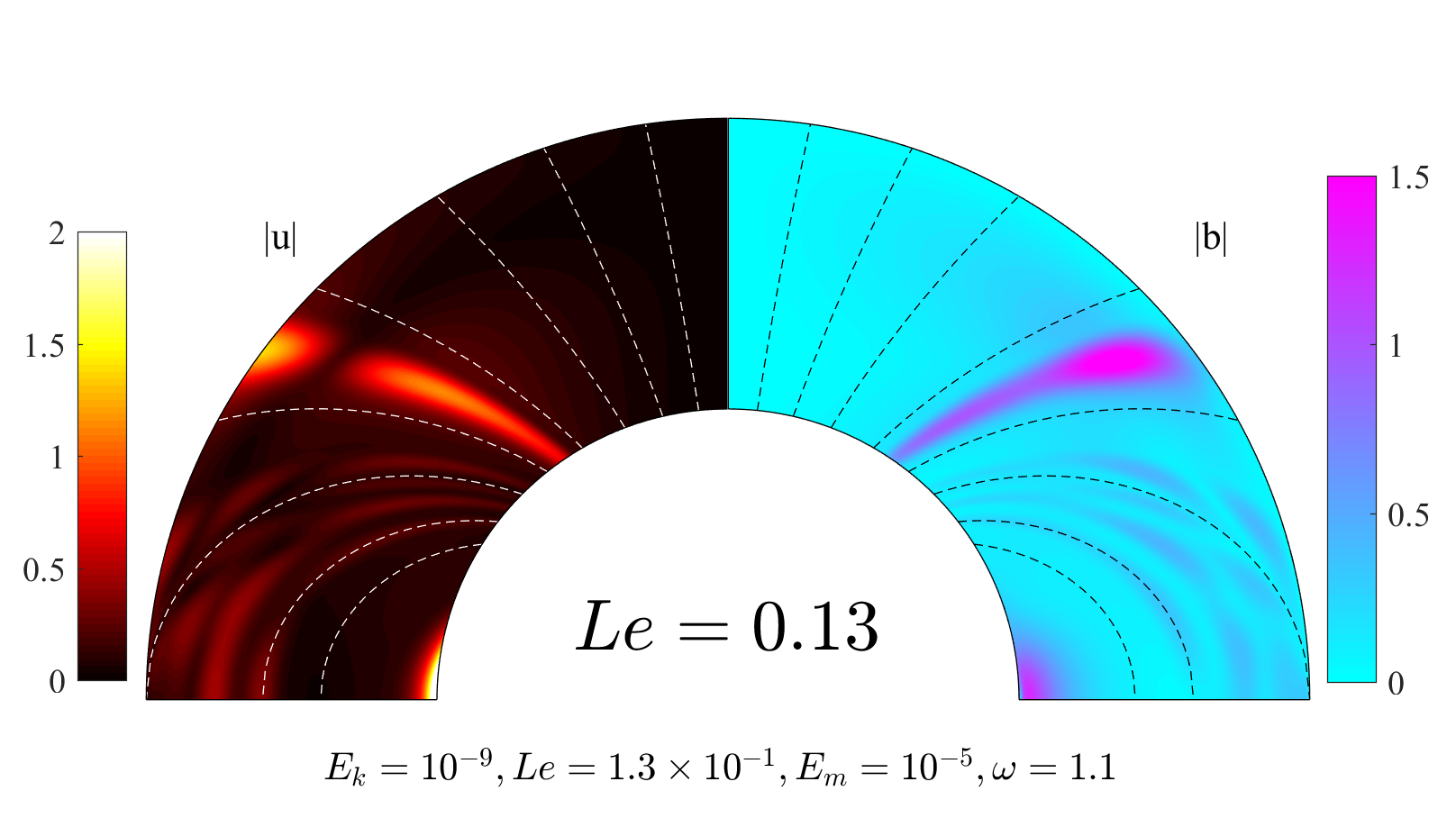}\\
(f)\includegraphics[width=0.48 \textwidth,trim={0cm 1.55cm 0cm 1.5cm},clip]{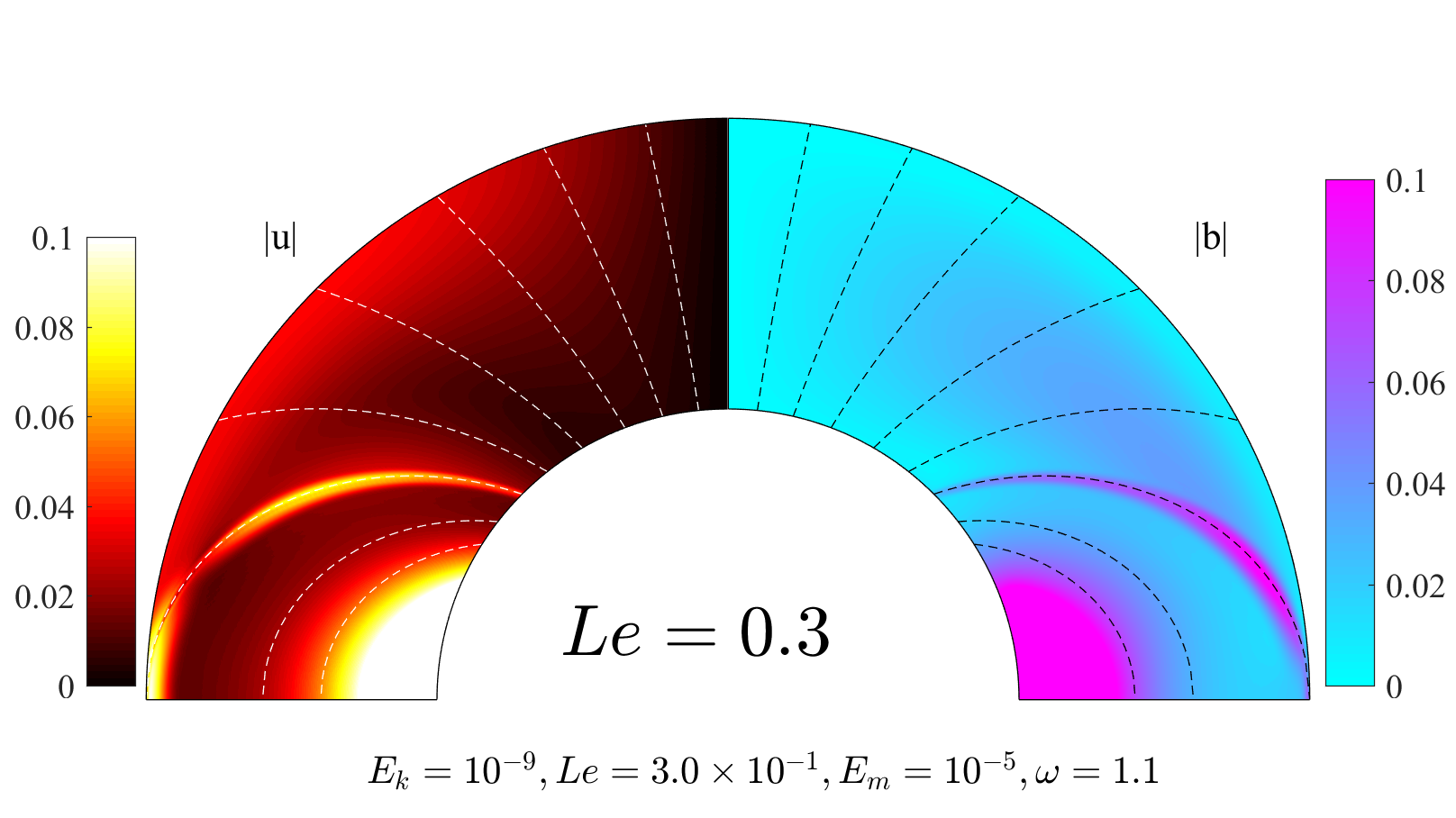} \\
\caption{Same as Fig. \ref{fig:Bz} but for a dipolar field $\bmath {B_0}$. Dashed lines show some field lines of $\bmath {B_0}$. }
\label{fig:dipole} 
\end{center}
\end{figure}

Fig. \ref{fig:dipole} shows the structure of the perturbations as in Fig. \ref{fig:Bz} but for a dipolar field $\bmath{B_0}$. The spatial structures vary in a similar way as in the case of an axial field, as we gradually increase $Le$. However, there are some local differences because the dipolar field $\bmath{B_0}$ is spatially non-uniform, being stronger at the polar regions than at the equator and decaying as a function of the radius. For instance, in Fig. \ref{fig:dipole}(c), the perturbations are less influenced by the magnetic field near the equator compared to other regions. Nevertheless, the general picture is qualitatively similar to Fig. \ref{fig:Bz}, from inertial wave attractors at small $Le$ to nearly Alfv\'en wave perturbations at relatively large $Le$.     

\begin{figure}
\begin{center}
\includegraphics[width=0.45\textwidth]{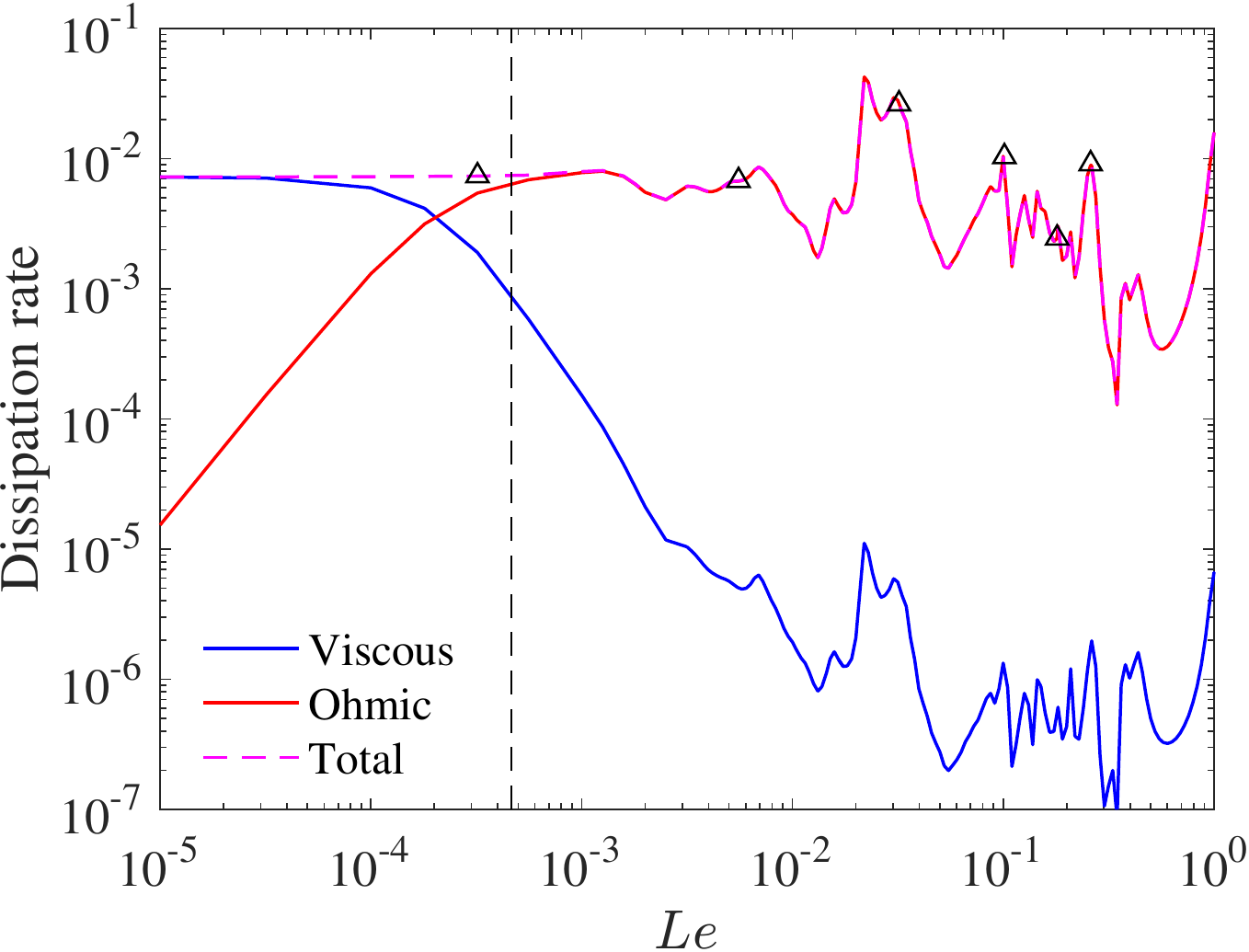} \\(a)
\includegraphics[width=0.45\textwidth]{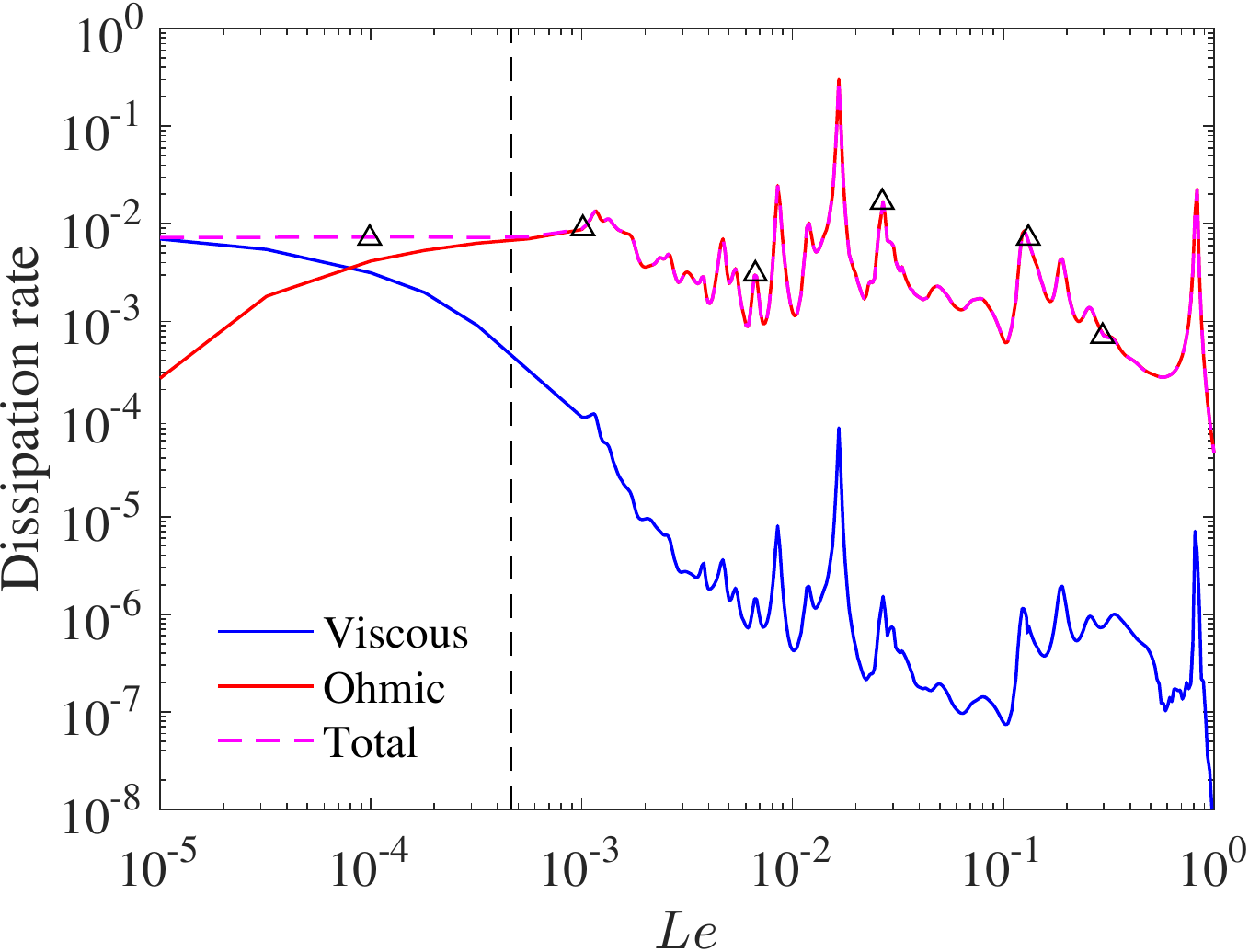} \\(b)\\
\caption{Dissipation rate versus the Lehnert number $Le$ for (a) an axial field, (b) a dipolar field. Vertical dash lines represent $Le=E_m^{2/3}$. $E_k=1.0\times 10^{-9}$, $E_m=1.0\times 10^{-5}$, $\omega=1.1$, $\alpha=0.5$. Black triangles correspond to cases shown in Figs. \ref{fig:Bz}-\ref{fig:dipole}.}
\label{fig:Diss_as_Le}
\end{center}
\end{figure}

We have mentioned that the perturbations retain the ray dynamics of inertial waves when $Le\le O(E_m^{2/3})$. We can derive this scaling by simply comparing several typical time scales in the system when $Le \ll 1$, $E_k\ll 1$, $E_m\ll 1$ and $Pm\ll 1$. The inertial wave propagation time in the fluid domain is 
\begin{equation} \label{eq:tau_i}
\tau_i=\frac{L}{|\bmath {V_g}|}\sim\left(\frac{l}{L}\right)^{-1} \Omega^{-1}.
\end{equation} 
The time scale for  Alfv\'en waves to transversely cross the inertial wave beams is
\begin{equation} \label{eq:tau_a}
\tau_a=\frac{l}{|\bmath{V_a}|}\sim \frac{l}{L}Le^{-1} \Omega^{-1}.
\end{equation}
The magnetic diffusion time across the beams is
\begin{equation} \label{eq:tau_eta}
\tau_{\eta}=\frac{l^2}{\eta}\sim \left(\frac{l}{L}\right)^{2}E_m^{-1} \Omega^{-1},
\end{equation}
and the viscous diffusion time is
\begin{equation}
\tau_{\nu}=\frac{l^2}{\nu}\sim \left(\frac{l}{L}\right)^{2}E_k^{-1} \Omega^{-1}.
\end{equation}
Here $l$ is the typical width of the wave beams, whereas $L$ is the domain size, i.e. $L=R$. The viscous time scale is irrelevant here because of the assumption $Pm\ll 1$. In order to keep the perturbations within the inertial wave beams, the crossing time of Alfv\'en waves $\tau_a$ should be longer than the inertial wave propagation time $\tau_i$:
\begin{equation} \label{eq:time1}
\tau_a\ge \tau_i.
\end{equation}
Meanwhile, the width of the wave beams is set by the diffusion time across the beams:
\begin{equation} \label{eq:time2}
\tau_i=\tau_\eta.
\end{equation}
Combining equations (\ref{eq:time1}-\ref{eq:time2}) and using equations (\ref{eq:tau_i}-\ref{eq:tau_eta}), we obtain
\begin{equation}\label{eq:scaling}
Le\le O(E_m^{2/3}).
\end{equation}
Note that this scaling analysis is merely heuristic. The prefactor of the scaling (\ref{eq:scaling}) varies depending on the frequency and the structure of background fields. Nevertheless, the above scaling provides an approximate threshold, above which the presence of a magnetic field would modify the propagation of inertial waves. This scaling is also clearly evidenced from the dissipation rate in  Fig. \ref{fig:Diss_as_Le}.

Fig. \ref{fig:Diss_as_Le} shows the viscous dissipation rate, the Ohmic dissipation rate and the total dissipation rate as a function of $Le$ for an axial field (a) and a dipolar field (b). When $Le \le O(E_m^{2/3})$, the dissipation rate due to the Ohmic damping grows and then saturates as $Le$ increases, while the viscous dissipation rate drops and eventually becomes negligible. However, the total dissipation rate remains unchanged in the range of $Le \le O(E_m^{2/3})$. This observation is reminiscent of the analytical result by \cite{Ogilvie2005JFM}, who has shown that the total rate of energy dissipation of a wave attractor is independent of the dissipative properties and the  detailed damping mechanisms. The theoretical analysis has been confirmed by hydrodynamic calculations in spherical shells \citep{Ogilvie2009MNRA,Rieutord2010}. Here we show that the theory is still valid  in the presence of a magnetic field as long as the wave attractors are retained. 
%Energy can be dissipated either by the viscosity or by the Ohmic heating, or both of them, but the total dissipation rate remains unchanged as increasing $Le$ or varying $E_k$ and $E_m$, provided that $Le \le O(E_m^{2/3})$.

When $Le>O(E_m^{2/3})$, the total dissipation is almost totally contributed by the Ohmic damping, whereas the viscous dissipation is negligible. The dissipation rate fluctuates as a function of $Le$, exhibiting several peaks and troughs. In this range of parameters, the magnetic field modifies the propagation of waves depending on the Lehnert number. Resonance may occur at certain values of $Le$ for a given frequency, leading to enhanced dissipation. Indeed, these peaks in the dissipation rate usually correspond to either large-scale perturbations, e.g. Fig \ref{fig:Bz}(d) and Fig \ref{fig:dipole} (d),  or perturbations concentrating on certain field lines, e.g. Fig \ref{fig:Bz} (f) and Fig \ref{fig:dipole} (e).  

Note that we have restricted our investigations in the range of $Le\le 1$. We found that strong magnetic boundary layers arise when $Le \gg 1$, and thus the energy dissipation is mainly contributed by the boundary layers, which may be not realistic because of our idealized boundary conditions. For instance, a rigid core of finite electric  conductivity (rather than insulating) would relax the accumulation of the electrical current near the inner boundary. Anyway, the Lehnert number should be smaller than unity in most stars and planets, although the specific value is difficult to estimate owing to the uncertainties of the magnetic field strength and fluid properties. The Lehnert number for the Sun is estimated to be around $10^{-5}$, if we assume the typical magnetic field strength  is a few $10^{-3}$~T \citep{Charbonneau2014ARAA}, and use the mean density of the Sun. Meanwhile, the magnetic Ekman number is also very small for the Sun, i.e. $E_m<10^{-10}$, as the magnetic diffusion time is around $10^{10}$ year \citep{Charbonneau2014ARAA}. The magnetic field is strong enough to modify the propagation of inertial waves as $Le>O(E_m^{2/3})$ for the Sun, despite the small Lehnert number.

\subsection{Frequency-dependence} \label{subsec:fre_scan}
\begin{figure*}
\includegraphics[width=0.95 \textwidth]{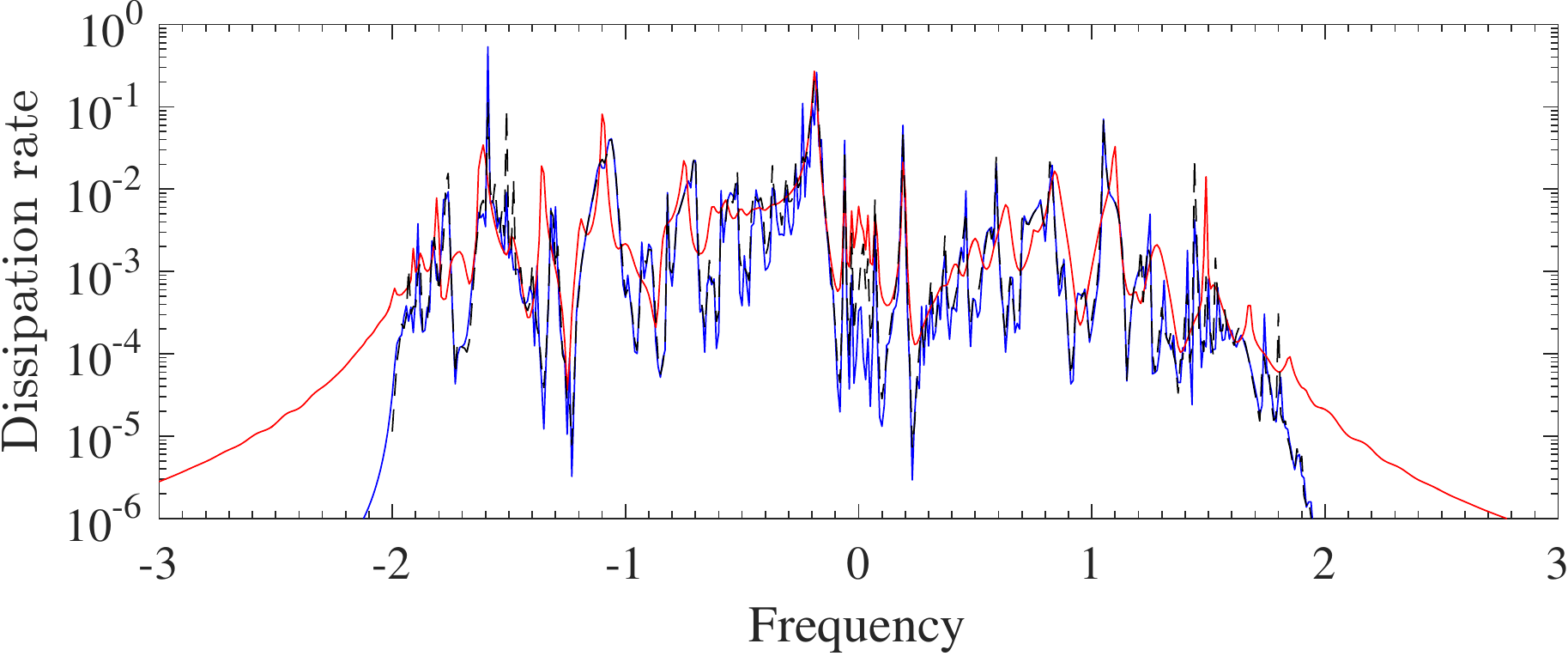}  \\
(a) \\
\includegraphics[width=0.95 \textwidth]{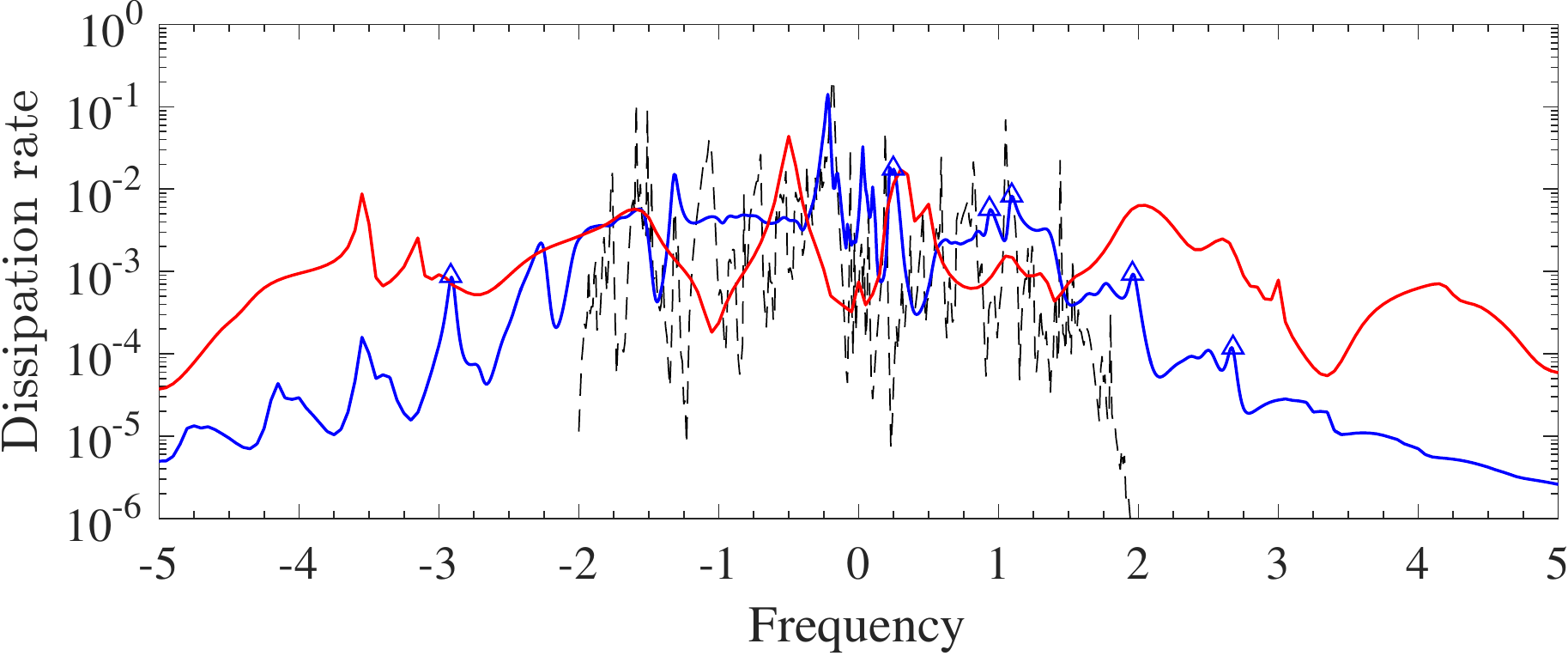} \\
(b)
\caption{Total dissipation rate versus the tidal forcing frequency. The background magnetic field is set to be an axial field. $\alpha=0.5$, $E_k=10^{-8}$ and $E_m=10^{-4}$. (a) $Le=3.2 \times 10^{-3}$ (blue) and $Le=2.4 \times 10^{-2}$ (red); (b) $Le=0.1$ (blue) and $Le=0.42$ (red). Black dashed lines represent the dissipation rate in the absence of a magnetic field. Blue triangles correspond to cases shown in Fig \ref{fig:MCmode}.}
\label{fig:Diss_as_fre_Bz}
\end{figure*}
We now investigate the frequency-dependence of the dissipation rate. Fig \ref{fig:Diss_as_fre_Bz} shows a frequency scan of the total dissipation rate in a frequency range of $-3\le \omega \le 3$ at various values of $Le$ with an axial field, and $\alpha=0.5$, $E_k=10^{-8}$ and $E_m=10^{-4}$. For comparison, we show also the dissipation rate in the absence of a magnetic field, i.e. $Le=0$, in the frequency range of inertial waves. In the presence of the magnetic field, we show only the cases when $Le>O(E_m^{2/3})$, because the total dissipation rate shows similar behaviour to that of inertial waves when $Le\le O(E_m^{2/3})$.

Fig \ref{fig:Diss_as_fre_Bz} (a) shows the dissipation rate at $Le=3.2 \times 10^{-3}$ (the blue curve) and $Le=2.4 \times 10^{-2}$ ( the red curve). We can see that these curves are very bumpy, likewise the curve in the absence of a magnetic field (the black dashed line). The peaks and troughs are closely related to those the black dashed line. These cases can be regarded as weakly modified inertial waves. In particular, we can see from the red curve that the peaks shift to higher frequencies, which is in line with the dispersion relation of magnetic-Coriolis waves. 

Fig \ref{fig:Diss_as_fre_Bz} (b) shows the dissipation rate at $Le=0.1$ (the blue curve) and $Le=0.42$ ( the red curve). These two curves are significantly smoothed out by the presence of a magnetic field compared to the black dashed line. In addition, the spectra of the dissipation rate are broadened beyond the frequency range of inertial waves, as expected from the dispersion relation (\ref{eq:dispersion_nd}).
\begin{figure}
\begin{center}
(a)\includegraphics[width=0.48 \textwidth,trim={0cm 1cm 0cm 1cm},clip]{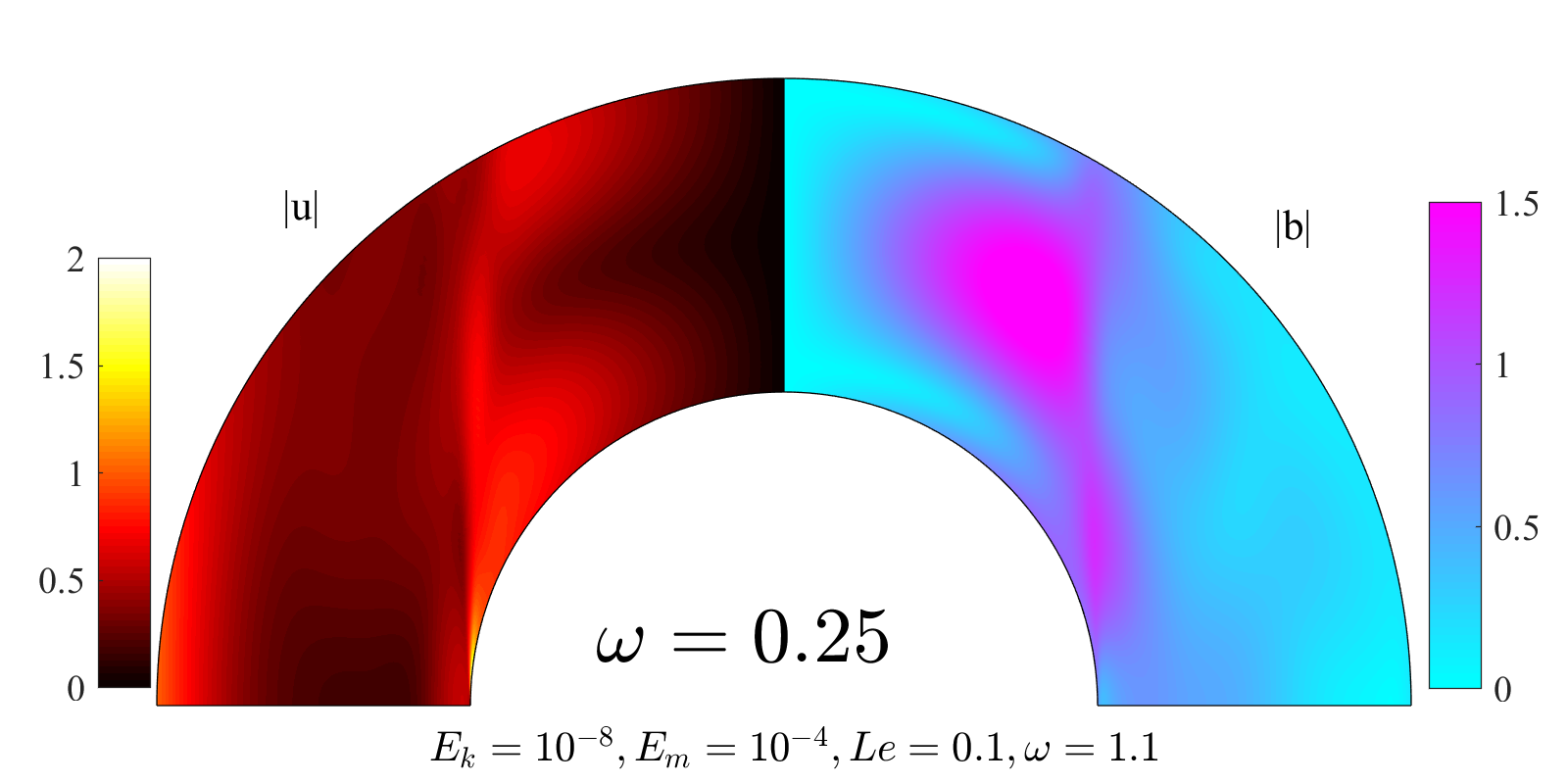}\\
(b)\includegraphics[width=0.48 \textwidth,trim={0cm 1cm 0cm 1cm},clip]{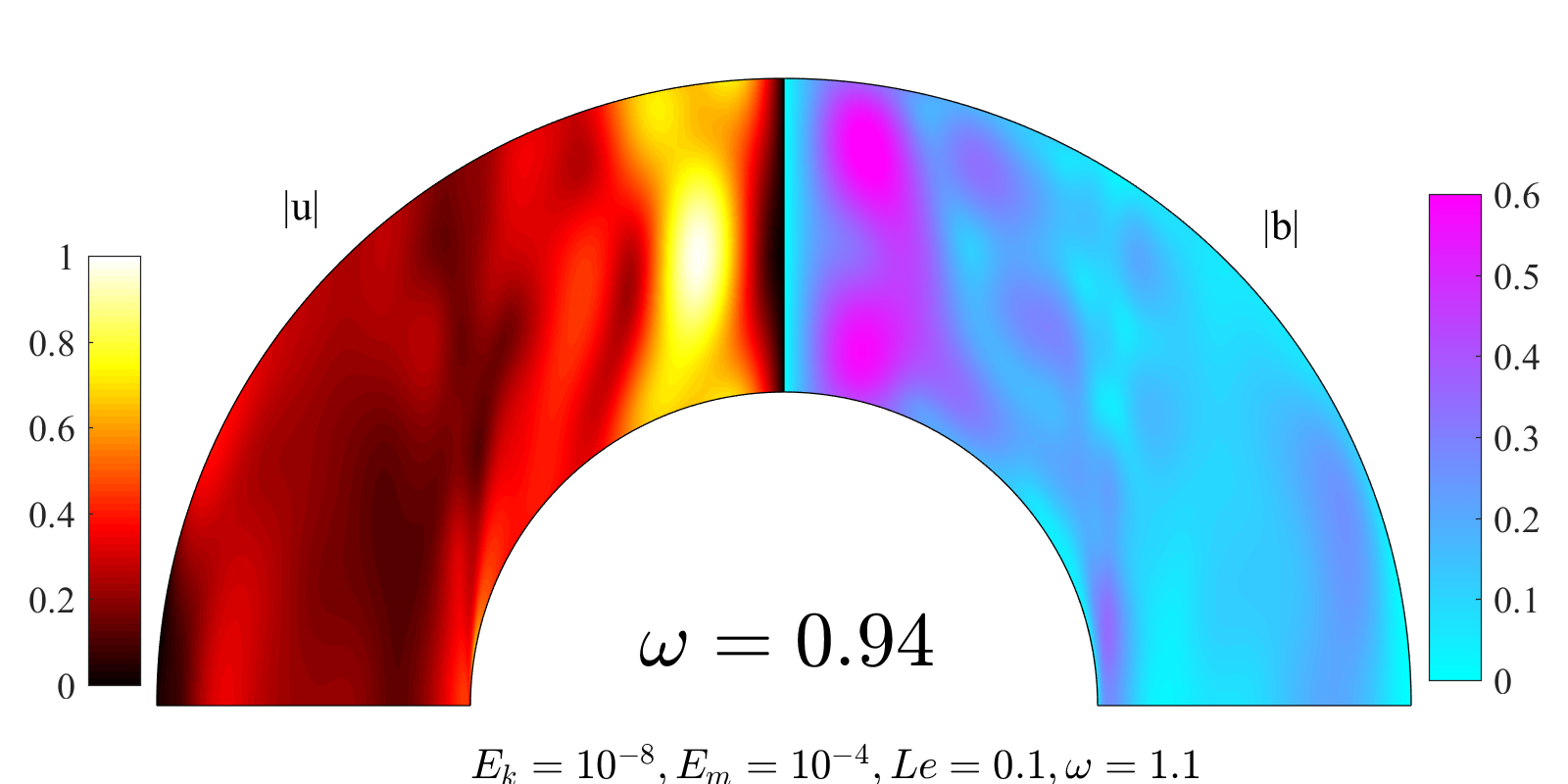}\\ 
(c)\includegraphics[width=0.48 \textwidth,trim={0cm 1cm 0cm 1cm},clip]{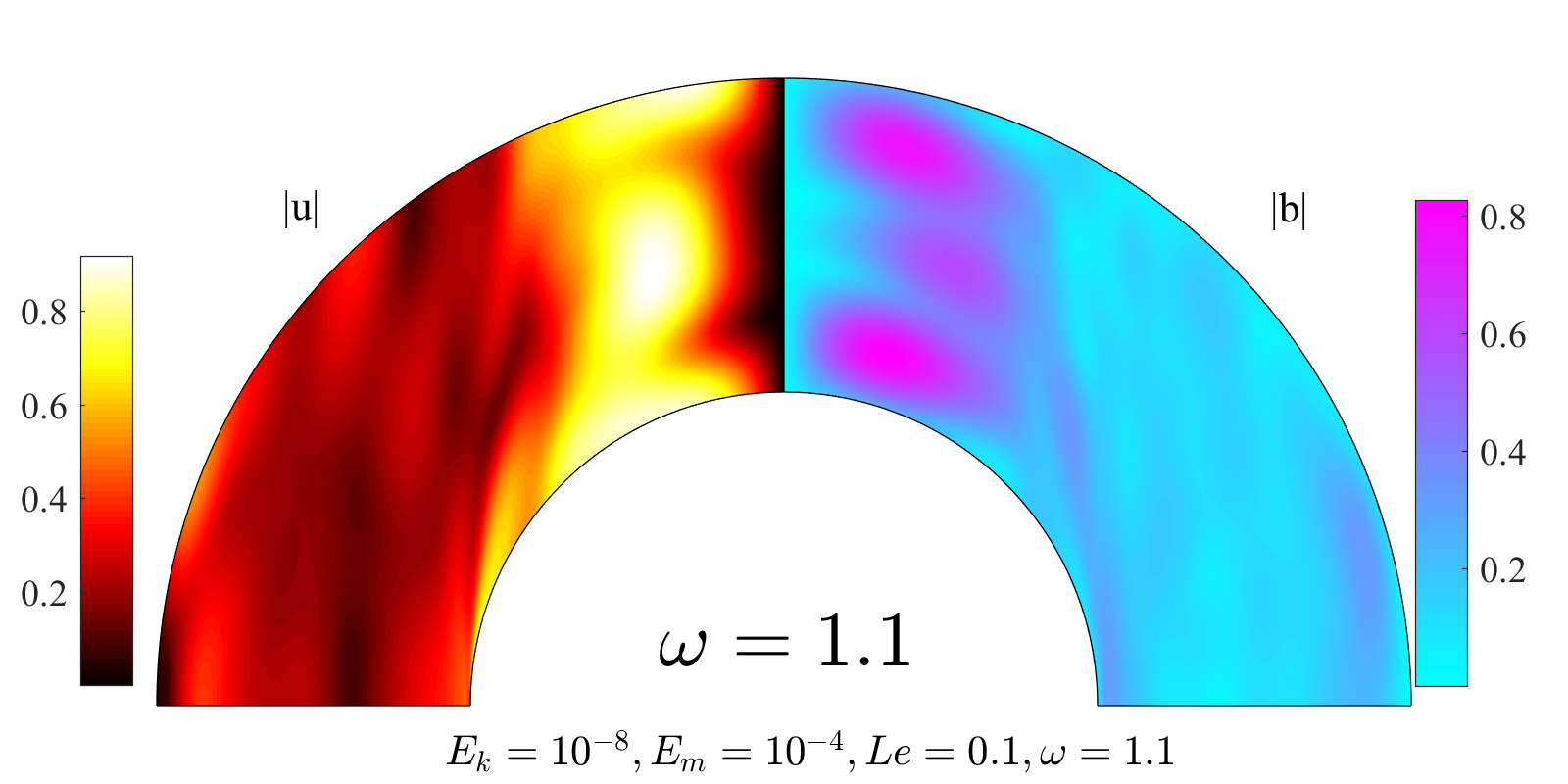}\\
(d)\includegraphics[width=0.48 \textwidth,trim={0cm 1cm 0cm 1cm},clip]{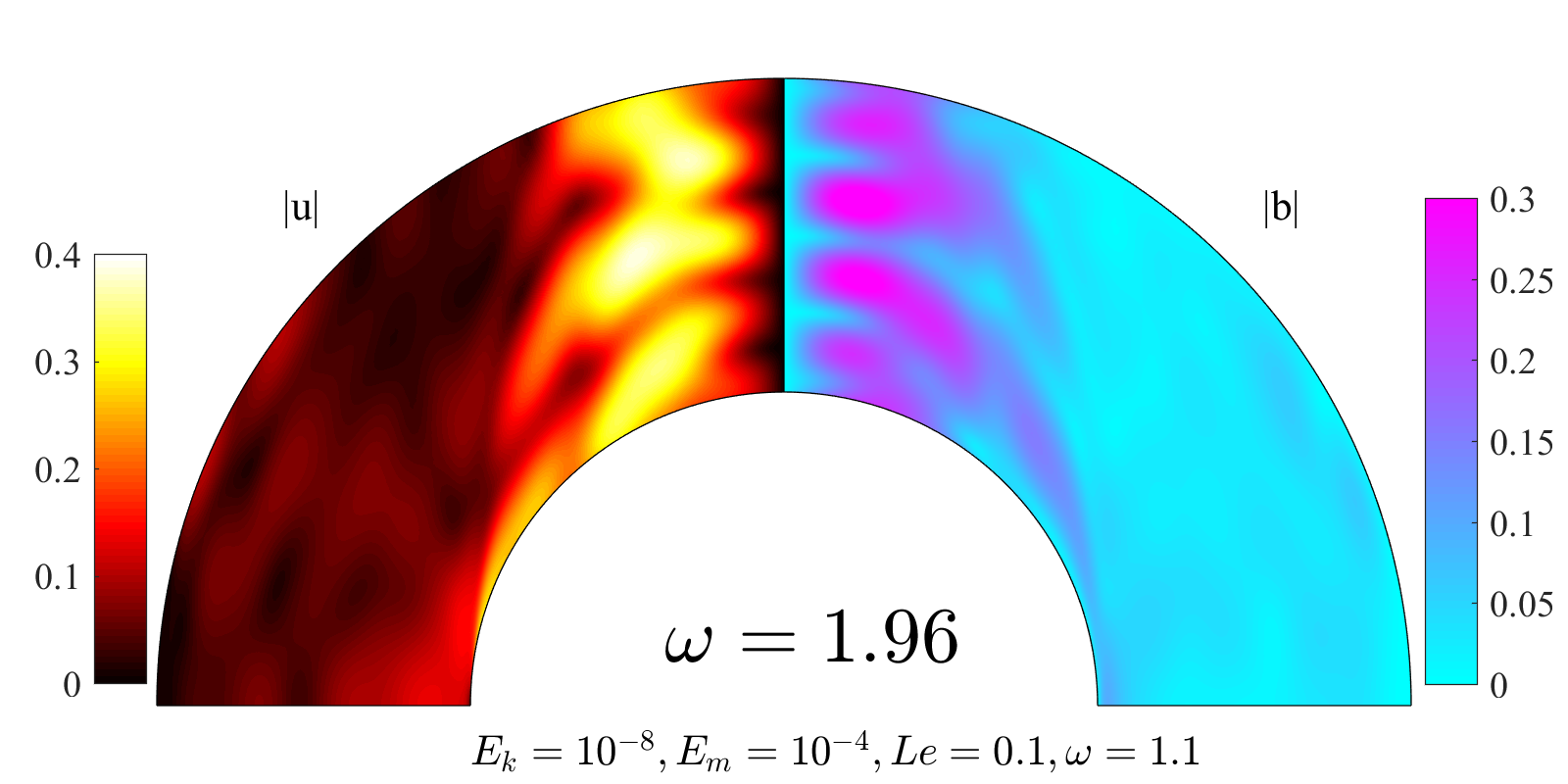}\\
(e)\includegraphics[width=0.48 \textwidth,trim={0cm 1cm 0cm 1cm},clip]{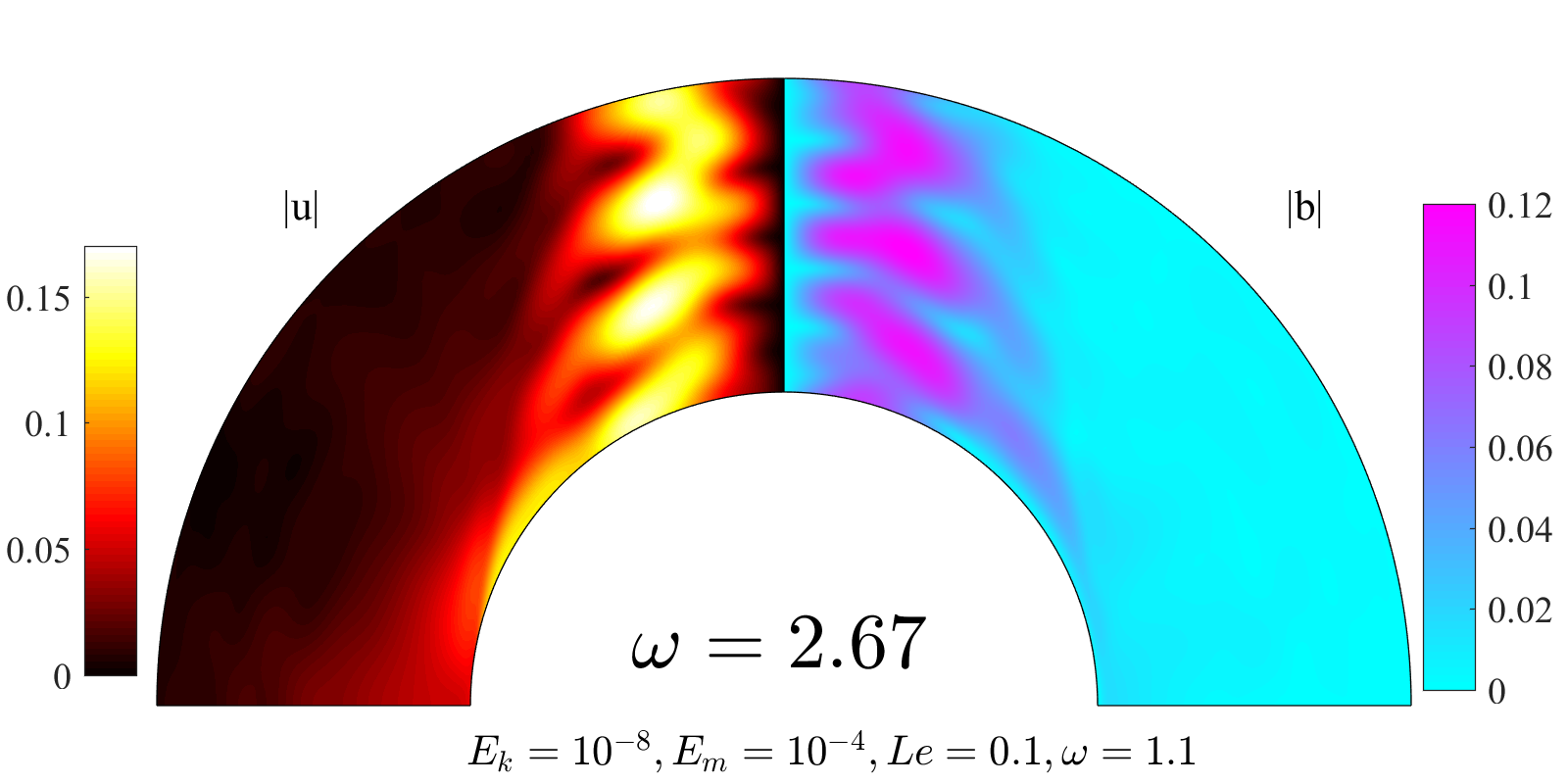}\\
(f)\includegraphics[width=0.48 \textwidth,trim={0cm 1cm 0cm 1cm},clip]{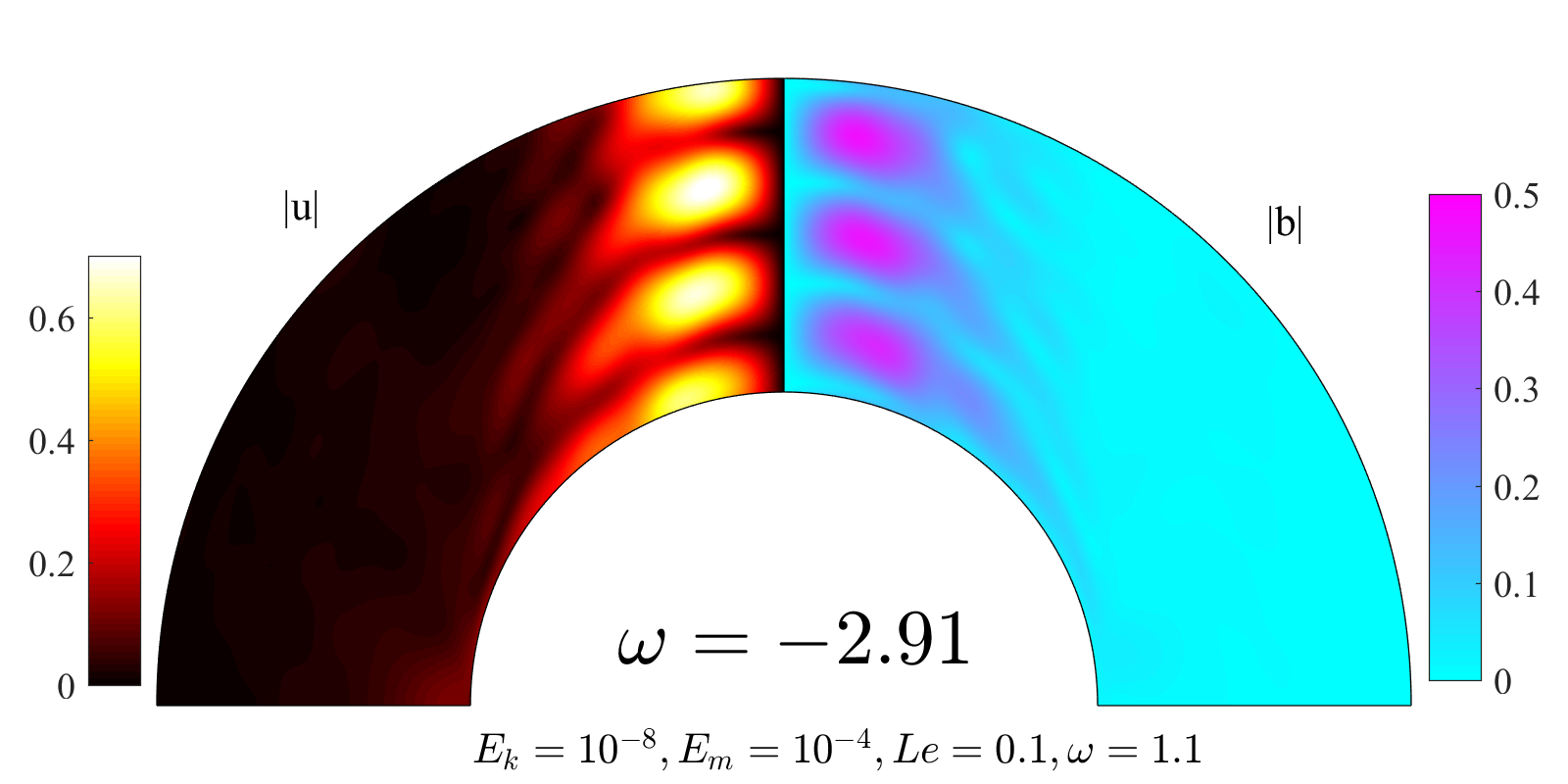} \\
\caption{Structure of the velocity perturbation $|\bmath u|$ and the magnetic field perturbation $|\bmath b|$ in the meridional plane with an axial field $\bmath{B_0}$ at different tidal frequencies. $Le=0.1$, $E_k=10^{-8}$, $E_m=10^{-4}$,  $\alpha=0.5$.} 
\label{fig:MCmode} 
\end{center}
\end{figure}

Although the dissipation rate curves become smooth at relatively large $Le$, they still exhibit a few peaks in the frequency range we have shown. Fig. \ref{fig:MCmode} shows structures of the perturbations at some peak frequencies of the blue curve in Fig. \ref{fig:Diss_as_fre_Bz} (b). We can see that all these cases feature smooth large-scale  structures, in particular in the polar region. We also note that the number of nodes in the vertical direction increases as the frequency increases, which is reminiscent of the dispersion relation of magnetic-Coriolis waves. We have mentioned that such smooth structures may be eigen-modes of the system, which are resonantly excited at the eigen-frequencies, leading to the enhanced energy dissipation. However, the theoretical analysis of the eigen value problem is beyond the scope of this paper.

For the case of a dipolar field, the frequency-dependence of the total dissipation rate is qualitatively similar to that of an axial field, although details are different, such as the peak frequencies.
 
\subsection{Frequency-averaged dissipation rate} \label{subsec:FreAce}
\begin{figure}
\includegraphics[width=0.5 \textwidth]{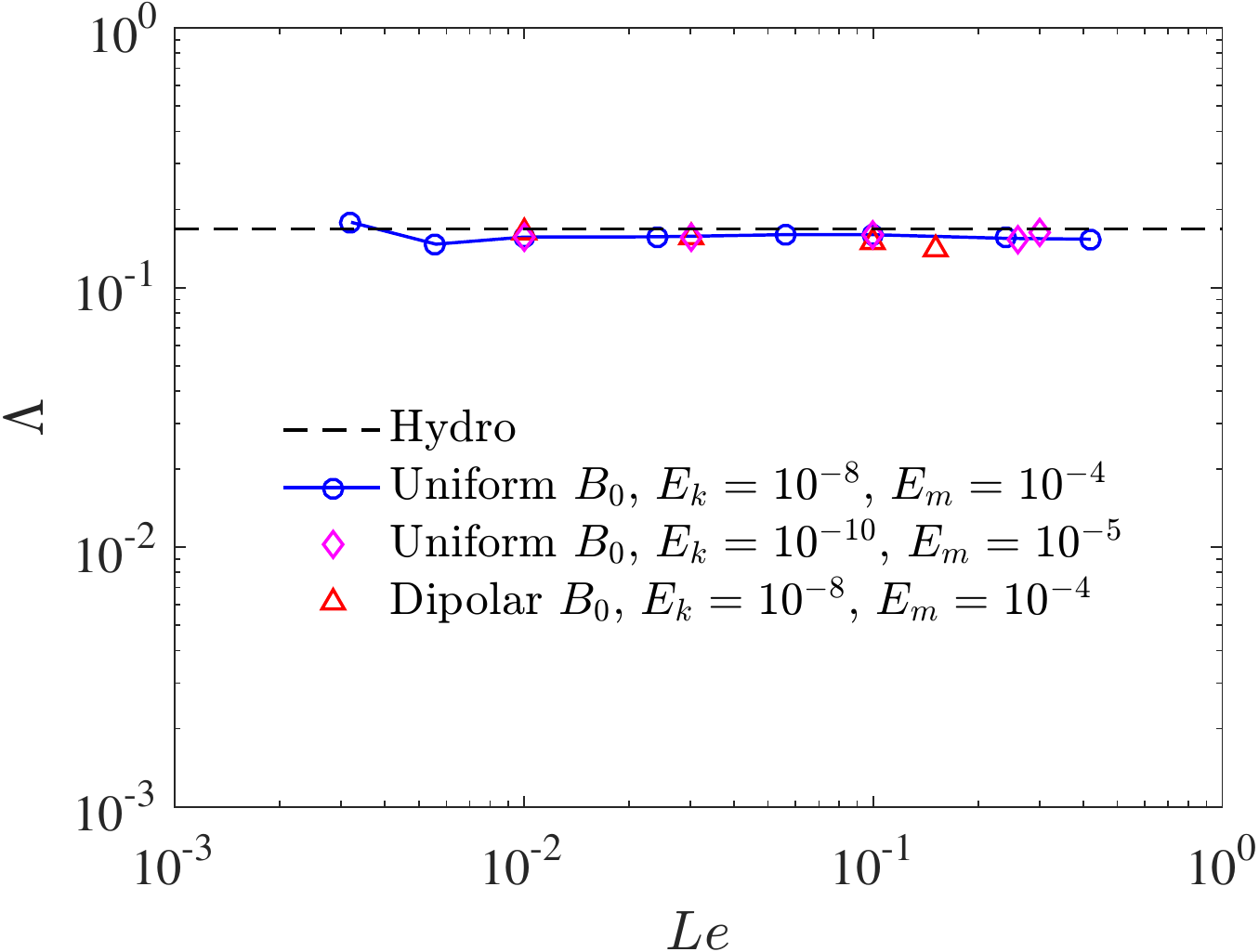}
\caption{Frequency-averaged quantity $\Lambda$ versus the Lehnert number $Le$ for various different parameters but for fixed inner core radius $\alpha=0.5$.}
\label{fig:Fre_ave_diss}
\end{figure}
The tidal dissipation leads to a long-term evolution of the spin and orbital parameters through the exchange of angular momentum. As the system evolves, the tidal frequency varies over time. It is very difficult to estimate the instantaneous tidal dissipation owing to the complicated frequency-dependence. However, the frequency-averaged dissipation rate can be useful to study the long-term evolution of the system \citep{Mathis2015AA,Bolmont2016CeMDA, Gallet2017,Bolmont2017}. \cite{Ogilvie2013MNRAS} has shown that the frequency-averaged dissipation rate of inertial waves is independent of the dissipative properties, but strongly depends on the size of the rigid core. Here we examine the frequency-averaged dissipation in the presence of a magnetic field. 

The frequency-averaged dissipation can be measured by the dimensionless quantity \citep{Ogilvie2013MNRAS}
\begin{equation}
\Lambda=\int_{-\infty}^{\infty}\mathrm{Im}[K_l^m(\omega)]\frac{\mathrm{d} \omega}{\omega},
\end{equation}
where $\mathrm{Im}[K_l^m]$ is the imaginary part of the potential Love number, and is related to the dissipation rate by 
\begin{equation}
\hat{D}=\frac{(2l+1)R}{8 \pi G}A^2\Omega\omega\mathrm{Im}[K_l^m(\omega)].
\end{equation}
where $A$ is the tidal amplitude with the unit of gravitational potential and $\hat{D}$ is the dimensional dissipation rate with the unit of power. With our normalization of the forcing in equation (\ref{eq:forcing_nd}) for $l=2$ and $m=2$, the frequency-averaged quantity becomes
  \begin{equation}
\Lambda=\int_{-\infty}^{\infty}\mathrm{Im}[K_2^2(\omega)]\frac{\mathrm{d} \omega}{\omega}=\frac{15}{2} \epsilon^2 \int_{-\infty}^{\infty}D(\omega)\mathrm{d} \omega,
\end{equation} 
where $\epsilon^2=\Omega^2R^3/GM$ and $D(\omega)=D_{vis}+D_{ohm}$ is the dimensionless total dissipation rate as shown in Fig. \ref{fig:Diss_as_fre_Bz}. Note that $\epsilon$ is a small parameter for astrophysical bodies, but we simply set $\epsilon=1$ in our linear calculations of the wavelike perturbations. The above integral can be carried out only in a finite frequency range numerically. For the hydrodynamic case ($Le=0$), the integral is evaluated over the frequency range of inertial waves, i.e. $-2<\omega<2$, as in \cite{Ogilvie2013MNRAS}. The spectrum of MC waves is unlimited according to the dispersion relation (\ref{eq:dispersion_nd}). However, the major dissipation still occurs in a finite frequency range in our calculations of $Le<1$. 
%The frequency range is broadened accordingly as increasing $Le$ by examining the dissipation curves. 
The integral is typically evaluated over the frequency range of  $-3<\omega<3$ when $Le\leq 0.1$, and $-5<\omega<5$ when $Le>0.1$. We also doubled the frequency range at large $Le$ and found that the results of the integral are converged. Fig. \ref{fig:Fre_ave_diss} shows the frequency-averaged quantity $\Lambda$ at several different parameters but with the fixed inner core size $\alpha=0.5$. We can see that the frequency-averaged dissipation rate is independent of the strength and the structure of the magnetic field, and the dissipative parameters $E_k$ and $E_m$. Similar results have been observed in a previous study using a periodic box \citep{Wei2016ApJ}.  Also, the frequency-averaged dissipation rate in the presence of a magnetic is nearly the same as that in the absence of a magnetic field. The small discrepancies are likely due to the errors of numerical integrals. %This suggests that the frequency-averaged dissipation rate is insensitive to detailed dissipation mechanisms for the wavelike part.

\begin{figure}
\includegraphics[width=0.5 \textwidth]{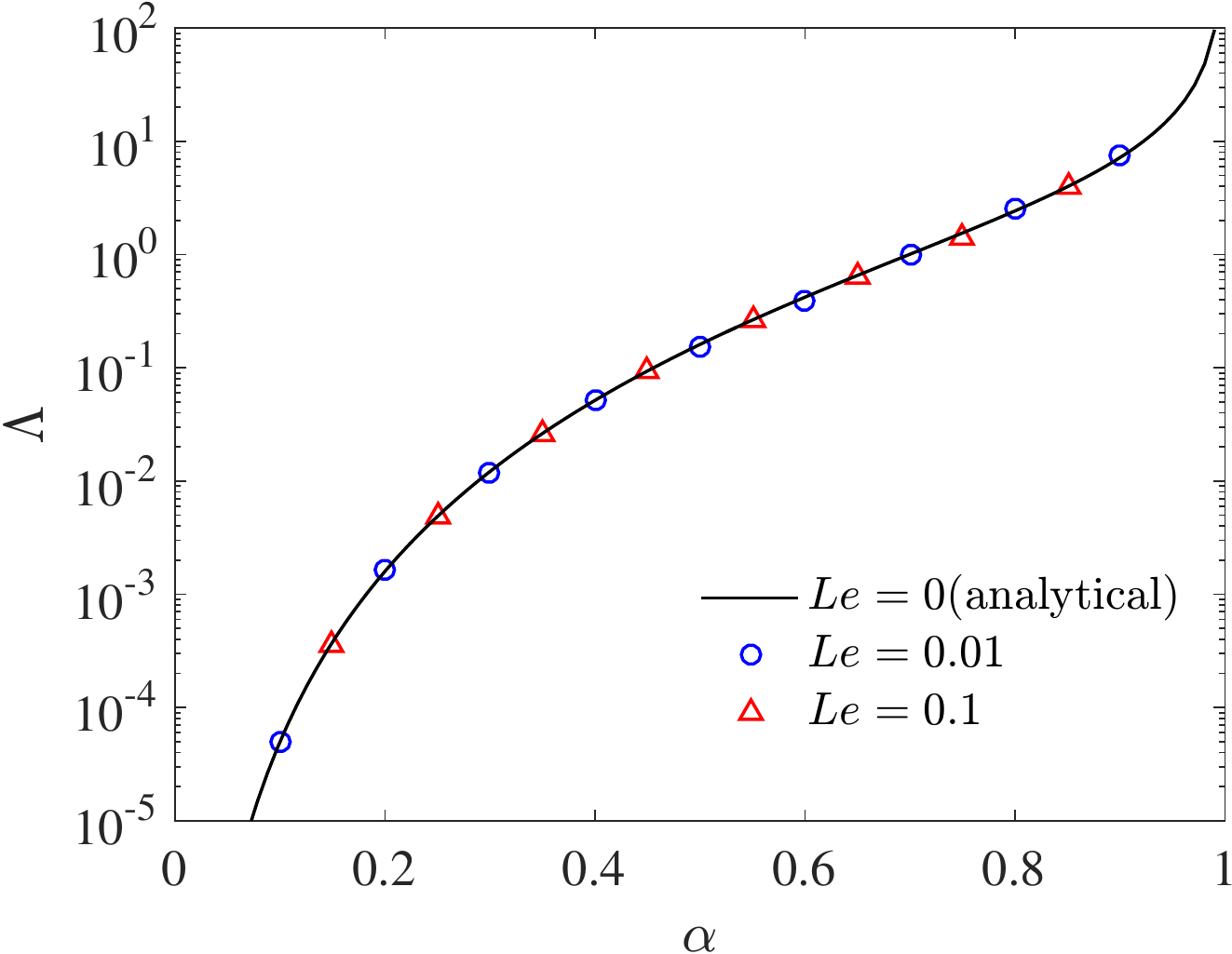}
\caption{Frequency-averaged quantity $\Lambda$ versus the radius ratio $\alpha$ at different values of $Le$ with an axial field and $E_k=10^{-8}$, $E_m=10^{-4}$. The solid line represents the analytical expression from Ogilvie (2013).}
\label{fig:Diss_as_ri}
\end{figure}

However, the frequency-averaged dissipation rate strongly depends on the size of the inner core. \cite{Ogilvie2013MNRAS} derived an analytical expression by considering low-frequency hydrodynamic responses to an impulsive forcing, which is given as
\begin{equation}\label{eq:Ogilvie2013}
\Lambda=\int_{-\infty}^{\infty}\mathrm{Im}[K_2^2(\omega)]\frac{\mathrm{d} \omega}{\omega}=\frac{100 \pi}{63}\epsilon^2\left(\frac{\alpha^5}{1-\alpha^5}\right),
\end{equation}  
for a homogeneous fluid of the same density as that of the rigid core, and for the tidal component of $l=m=2$. Equation (\ref{eq:Ogilvie2013}) has been verified numerically \citep{Ogilvie2013MNRAS}, and resembles the scaling of $\alpha^5$ for the small inner core size found by previous hydrodynamic studies \citep{Goodman2009ApJ,Ogilvie2009MNRA,Rieutord2010}.

Fig. \ref{fig:Diss_as_ri} show the frequency-averaged quantity $\Lambda$ as a function of the radius ratio $\alpha$ at different values of $Le$. Remarkably, this frequency-averaged quantity in the presence of a magnetic field is still in very good agreement with equation (\ref{eq:Ogilvie2013}), which is derived in the absence of magnetic fields. In the presence of a magnetic field, the energy is mainly dissipated through the Ohmic damping of magnetic-Coriolis waves, which occur over a wider frequency range, but the frequency-averaged dissipation rate is the same as that of viscous dissipation of inertial waves. This suggests that the frequency-averaged dissipation rate is independent of the detailed damping mechanisms. Indeed, we show in Appendix \ref{App:FreAve} that the analytical results on the frequency-averaged dissipation in \cite{Ogilvie2013MNRAS} are not altered by the presence of a magnetic field, in the approximation that the wave-like response is driven only by the Coriolis force acting on the non-wavelike tidal flow, if we assume that the frequencies of MC waves are small compared to those of acoustic and surface gravity waves.  This is the case in our numerical calculations of the Lehnert number $Le\leq 1$ (see Fig. \ref{fig:Diss_as_fre_Bz}). In real astrophysical fluid bodies, this assumption may be justified by the fact the tidal frequency is usually small compared to the frequencies of acoustic and surface gravity waves.

\subsection{Obliquity tide}
\begin{figure}
\begin{center}
\includegraphics[width=0.5 \textwidth]{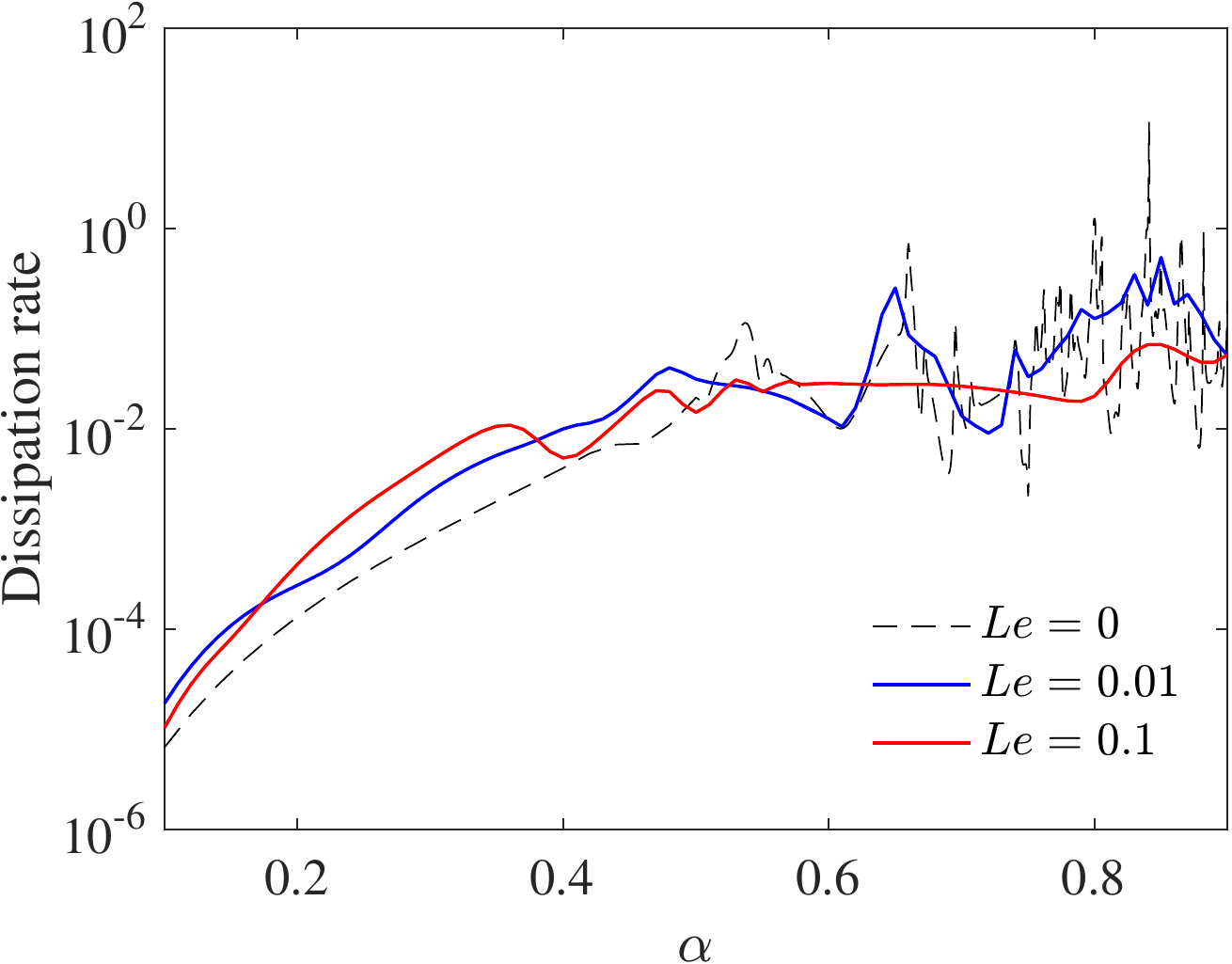}
\caption{Total dissipation rate of the obliquity tide as a function of radius ratio $\alpha$ at different values of $Le$ with an axial field. $E_k=10^{-7}$, $E_m=10^{-4}$.} \label{fig:Obliquity}
\end{center}
\end{figure}
So far, we considered only the tidal component of $l=2$ and $m=2$, which mainly determines the orbital evolution and synchronization. In this section, we briefly consider another important tidal component of $l=2$ and $m=1$, the so-called obliquity tide, which exists only in spin-orbit misaligned systems and mainly determines the evolution of the spin-orbit angle. The obliquity tide is peculiar because the tidal frequency in the rotating frame is always equal to $-\Omega$ regardless of the orbital frequency, i.e. $\omega=-1$ in dimensionless form. In addition, the obliquity tide is responsible for the precessional motion of the spin axis and the orbital normal around the total angular momentum vector. We have shown that dissipative inertial waves can be excited by the obliquity tide on top of precession in a hydrodynamic study \citep{Lin2017MNRAS}. To examine the magnetic effect on the wavelike responses of the obliquity tide, we need to replace the forcing in equation (\ref{eq:forcing}) by
\begin{equation}
\bmath f = 2 \bmath{\Omega} \btimes \grad[X_l(r)Y_2^1(\theta,\phi)]\mathrm{e}^{\mathrm{i}\Omega t}+(\bmath{\Omega}\btimes \bmath{\Omega_p})\btimes \bmath{r},
\end{equation}
where the last term arises from the precessional motion around the total angular momentum vector. The precession frequency is determined by the tidal amplitude and given as \citep{Lin2017MNRAS}
\begin{equation}
\Omega_p=-\frac{15}{8}\sqrt{\frac{5}{6 \pi}}\frac{ R^3\Omega A}{(1-\alpha^5)GM\sin i},
\end{equation} 
where $i$ is the angle between the spin angular momentum and the total angular momentum.

Fig. \ref{fig:Obliquity} shows the total dissipation as a function of the radius ratio $\alpha$ at different value of $Le$ with an axial magnetic field. For comparison, we show also the dissipation rate in the absence of a magnetic field, which exhibits complicated dependence of the core size owing to varied ray dynamics of inertial waves, especially when $\alpha>0.5$ \citep{Lin2017MNRAS}. The major effect of a magnetic field  is, again, to smooth out the dissipation rate curves, but the overall level is similar to that of the hydrodynamic case. Note that the frequency of the obliquity tide is always $\omega=-1$ in the rotating frame, so we did not explore the frequency-dependence and the frequency-averaged quantity for the obliquity tide.

\section{Conclusions}\label{sec:Conclusion}
We have investigated the magnetic effects on the tidal dissipation in rotating fluid bodies using a simplified model. The tidal responses are decomposed into the non-wavelike and wavelike parts, but we have focused on the latter in this study, which is in the form of magnetic-Coriolis waves. The linearized wave equations are numerically solved using a pseudo-spectral method. The major effects of the presence of a magnetic field can be summarized  as follows.

(i) When the magnetic field is very weak, namely $Le\leq O(E_m^{2/3})$, the wavelike perturbations retain the ray dynamics of inertial waves, while the energy can be dissipated through the viscous damping and the Ohmic damping. When $Le > O(E_m^{2/3})$, the magnetic field start to modify the propagation of waves, and the dissipation of energy occurs mainly through the Ohmic damping.

(ii) The magnetic field smooths out the complicated dependence of the total dissipation rate on the tidal frequency, and broadens the frequency spectrum of the dissipation rate, depending on the Lehnert number $Le$. 

(iii) However, the frequency-averaged dissipation quantity is independent of the magnetic field strength, the field structure and the dissipative parameters, but increases as the relative size of the rigid core in our simplified model. In more realistic models, this frequency-averaged quantity may depend on other properties of the internal structure such as the density profile. Indeed, the frequency-averaged quantity is in very good agreement with previous analytical results in the absence of magnetic fields \citep{Ogilvie2013MNRAS}. In Appendix \ref{App:FreAve}, we show that the magnetic field has no effect on the frequency-averaged dissipation, in the approximation that the wave-like response is driven only by the Coriolis force acting on the non-wavelike tidal flow, if we assume that the frequencies of magnetic-Coriolis waves are small compared to those of acoustic and surface gravity waves.

Owing to the complicated frequency-dependence of the dissipation rate of inertial waves, it is very difficult to directly apply the instantaneous tidal dissipation to astrophysical bodies. Alternatively, the frequency-averaged dissipation rate has been used to study tidal dissipation and evolution of stars \citep{Mathis2015AA,Bolmont2016CeMDA, Gallet2017,Bolmont2017}. It has been conjectured that other effects such as non-linear interactions, convection, differential rotations and magnetic fields may wash out some of the complicated frequency-dependence of purely inertial waves \citep{Ogilvie2013MNRAS}. Indeed, our numerical results show a smoothing effect of the magnetic field on the frequency-dependence. More complicated processes could lead to smoother dissipation curves. Therefore, the frequency-averaged dissipation quantity is probably a useful indicator of the efficiency of tidal dissipation. In addition, our results suggest that the frequency-averaged quantity is insensitive to the detailed damping mechanisms and dissipative properties. If this is still the case in more realistic models, it would be very useful in applications as these details are not well understood in stars and planets. Note that the frequency-averaged dissipation does depend on the internal structure of the bodies, which is merely determined by the size of the rigid core in our simplified model.

It is worthwhile to mention that we considered only the magnetic effects on the wave-like tidal perturbations due to the Coriolis force in this study. The magnetic field can also interact with large-scale non-wave-like motions to produce a further wave-like response, which remains to be studied. Our tentative investigations based on a radially forced model \citep{Ogilvie2009MNRA} suggest that the interactions between a magnetic field and the non-wavelike motions may be not negligible when the Lehnert number $Le>0.1$. However, the boundary conditions probably need to be treated more carefully with a magnetic field, as the non-wavelike part is associated with the instantaneous tidal deformation. 

\section*{Acknowledgements}
We would like to thank the anonymous referee for a set of detailed and constructive comments, which helped improve the paper. YL acknowledges the support of the Swiss National Science Foundation through an advanced Postdoc.Mobility fellowship. 
%%%%%%%%%%%%%%%%%%%%%%%%%%%%%%%%%%%%%%%%%%%%%%%%%%



%\bibliographystyle{mnras}
\bibliographystyle{mnras}
\bibliography{MCwave} % if your bibtex file is called example.bib

%%%%%%%%%%%%%%%%%%%%%%%%%%%%%%%%%%%%%%%%%%%%%%%%%%

%%%%%%%%%%%%%%%%% APPENDICES %%%%%%%%%%%%%%%%%%%%%

\appendix
\section{Equations projected on to spherical harmonics}\label{App:Projected}
Taking the curl and the curl of the curl of equation (\ref{eq:Nod_Ns}), taking the curl of equation (\ref{eq:Nod_Indu}) and using the orthogonality of the vector spherical harmonics \citep{Rieutord1987GAFD}, we obtain the following projected equations. Similar equations were given in the appendix of \cite{Rincon2003AA}, but without the Coriolis term. The projection of the Coriolis force can be found in, e.g.  \citet{Rieutord1987GAFD}. The projected equations are given as
\begin{multline} \label{eq:A1}
-\mathrm{i}\omega_l \mathcal{D}_l(u_l^m)=-2\beta_l^m\mathcal{D}_{l-1}^w  (w_{l-1}^m)-2\beta_{l+1}^m \mathcal{D}_{l+1}^w (w_{l+1}^m) \\
+Le^2 \left[\mathrm{i} m\mathcal{L}_l^c(c_l^m)+\alpha_{l}^m\mathcal{L}_{l-1}^a (a_{l-1}^m)+\alpha_{l+1}^m\mathcal{L}_{l+1}^a (a_{l+1}^m) \right] \\+E \mathcal{D}_l^u(u_l^m)+f_l,
\end{multline}
\begin{multline} \label{eq:A2}
-\mathrm{i}\omega_l w_l^m=-2\alpha_l^m\mathcal{D}_{l-1}^u  (u_{l-1}^m)-2\alpha_{l+1}^m \mathcal{D}_{l+1}^u (u_{l+1}^m) \\
+Le^2 \left[\mathrm{i} m\mathcal{L}_l^a(a_l^m)+\alpha_{l}^m\mathcal{L}_{l-1}^c (c_{l-1}^m)+\alpha_{l+1}^m\mathcal{L}_{l+1}^c (c_{l+1}^m) \right] \\
+E \mathcal{D}_l^w(w_l^m)+f_{l-1}+f_{l+1},
\end{multline}
\begin{multline}\label{eq:A3}
-\mathrm{i}\omega a_l^m=\mathrm{i} m\mathcal{L}_l^w(w_l^m)+\alpha_{l}^m\mathcal{L}_{l-1}^u(u_{l-1}^m)+\alpha_{l+1}^m\mathcal{L}_{l+1}^u(u_{l+1}^m)\\+E_m\mathcal{D}_l^a(a_l^m),
\end{multline}
\begin{multline}\label{eq:A4}
-\mathrm{i}\omega c_l^m= \mathrm{i} m\mathcal{L}_l^u(u_l^m)+\alpha_{l}^m \mathcal{L}_{l-1}^w(w_{l-1}^m)+\alpha_{l+1}^m\mathcal{L}_{l+1}^w(w_{l+1}^m)\\
+E_m\mathcal{D}_l^c(c_l^m),
\end{multline}
where $\mathcal{D}$ and $\mathcal{L}$ denote linear differential operators:
\begin{equation}
\mathcal{D}_l=r\frac{\mathrm{d}^2}{\mathrm{d}r^2}+4\frac{\mathrm{d}}{\mathrm{d}r}-(l^2+l-2)\frac{1}{r}, 
\end{equation}
\begin{equation}
\mathcal{D}_{l-1}^w=\frac{\mathrm{d}}{\mathrm{d}r}-(l-1)\frac{1}{r},
\end{equation}
\begin{equation}
\mathcal{D}_{l+1}^w=\frac{\mathrm{d}}{\mathrm{d}r}+(l+2)\frac{1}{r}
\end{equation}
\begin{multline}
\mathcal{D}_l^u=r\frac{\mathrm{d}^4}{\mathrm{d}r^4}+8\frac{\mathrm{d}^3}{\mathrm{d}r^3}-2(l_p-6)\frac{1}{r}\frac{\mathrm{d}^2}{\mathrm{d}r^2} \\
-4l_p\frac{1}{r^2}\frac{\mathrm{d}}{\mathrm{d}r}+l_p(l_p-2)\frac{1}{r^3},
\end{multline}
\begin{multline}
\mathcal{L}_{l}^c=B_r\frac{\mathrm{d}^2}{\mathrm{d}r^2}+\left(\frac{2 B_r}{r}+\frac{\mathrm{d}B_r}{\mathrm{d}r}\right)\frac{\mathrm{d}}{\mathrm{d}r} \\
+\frac{\mathrm{d}B_r}{\mathrm{d}r}-l_p\frac{\mathrm{d}}{\mathrm{d}r}\left(\frac{B_\theta}{r}\right),
\end{multline}
\begin{multline}
\mathcal{L}_{l-1}^a=l_p\left[rB_r\frac{\mathrm{d}^3}{\mathrm{d}r^3}+\left(l B_\theta+r\frac{\mathrm{d}B_r}{\mathrm{d}r}+6B_r\right)\frac{\mathrm{d}^2}{\mathrm{d}r^2}\right] \\
-l_p\left[(l+2)(l-3)\frac{B_r}{r}-4\frac{\mathrm{d}B_r}{\mathrm{d}r}-4l\frac{B_\theta}{r}\right]\frac{\mathrm{d}}{\mathrm{d}r} \\
-l_p(l+1)(l-2)\left(l \frac{B_\theta}{r^2}+\frac{1}{r}\frac{\mathrm{d}B_r}{\mathrm{d}r}\right),
\end{multline}
\begin{multline}
\mathcal{L}_{l+1}^a=l_p\left[rB_r\frac{\mathrm{d}^3}{\mathrm{d}r^3}-\left((l+1)B_\theta-r\frac{\mathrm{d}B_r}{\mathrm{d}r}-6B_r\right)\frac{\mathrm{d}^2}{\mathrm{d}r^2}\right] \\
-l_p\left[(l+4)(l-1)\frac{B_r}{r}-4\frac{\mathrm{d}B_r}{\mathrm{d}r}-4(l+1)\frac{B_\theta}{r}\right]\frac{\mathrm{d}}{\mathrm{d}r} \\
+l_pl(l+3)\left((l+1) \frac{B_\theta}{r^2}-\frac{1}{r}\frac{\mathrm{d}B_r}{\mathrm{d}r}\right),
\end{multline}

\begin{equation}
\mathcal{D}_{l-1}^u=r\frac{\mathrm{d}}{\mathrm{d}r}-(l-2),
\end{equation}
\begin{equation}
\mathcal{D}_{l+1}^u=r\frac{\mathrm{d}}{\mathrm{d}r}+(l+3),
\end{equation}
\begin{equation}
\mathcal{D}_l^w=\frac{\mathrm{d}^2}{\mathrm{d}r^2}+\frac{2}{r}\frac{\mathrm{d}}{\mathrm{d}r}-l_p\frac{1}{r^2},
\end{equation}

\begin{equation}
\mathcal{L}_{l}^a=-\frac{B_r}{l_p^2}\left(r\frac{\mathrm{d}^2}{\mathrm{d}r^2}+4\frac{\mathrm{d}}{\mathrm{d}r}-(l_p-2)\frac{1}{r} \right),
\end{equation}
\begin{equation}
\mathcal{L}_{l-1}^c=l(l-1)\left(B_r\frac{\mathrm{d}}{\mathrm{d}r}+\frac{B_r}{r}+l\frac{B_\theta}{r}\right),
\end{equation}
\begin{equation}
\mathcal{L}_{l+1}^c=(l+1)(l+2)\left(B_r\frac{\mathrm{d}}{\mathrm{d}r}+\frac{B_r}{r}-(l+1)\frac{B_\theta}{r}\right),
\end{equation}

\begin{equation}
\mathcal{D}_l^a=\frac{\mathrm{d}^2}{\mathrm{d}r^2}+\frac{4}{r}\frac{\mathrm{d}}{\mathrm{d}r}+(2-l_p)\frac{1}{r^2},
\end{equation}
\begin{equation}
\mathcal{L}_l^w=\frac{B_r}{r},
\end{equation}
\begin{equation}
\mathcal{L}_{l-1}^u=l_p\left(B_r\frac{\mathrm{d}}{\mathrm{d}r}+2\frac{B_r}{r}+l\frac{B_\theta}{r} \right),
\end{equation}
\begin{equation}
\mathcal{L}_{l+1}^u=l_p\left(B_r\frac{\mathrm{d}}{\mathrm{d}r}+2\frac{B_r}{r}-(l+1)\frac{B_\theta}{r} \right),
\end{equation}
\begin{equation}
l_p=l(l+1)
\end{equation}

\begin{equation}
\mathcal{D}_l^c=\frac{\mathrm{d}^2}{\mathrm{d}r^2}+\frac{2}{r}\frac{\mathrm{d}}{\mathrm{d}r}+l_p\frac{1}{r^2},
\end{equation}

\begin{multline}
\mathcal{L}_l^u=-\frac{1}{l_p^2}\left[ rB_r \frac{\mathrm{d}^2}{\mathrm{d}r^2}+\left(4B_r+r \frac{\mathrm{d}B_r}{\mathrm{d}r} \right)\frac{\mathrm{d}}{\mathrm{d}r}\right] \\
\frac{1}{l_p^2}\left(l_p \frac{\mathrm{d}B_\theta}{\mathrm{d}r}-l_p \frac{B_\theta}{r}-\frac{\mathrm{d}B_r}{\mathrm{d}r}-\frac{B_r}{r} \right),
\end{multline}
\begin{equation}
\mathcal{L}_{l-1}^w=l(l-1)\left(B_r \frac{\mathrm{d}}{\mathrm{d}r}+\frac{B_r}{r}+\frac{\mathrm{d B_r}}{\mathrm{d}r} +l\frac{B_\theta}{r}\right),
\end{equation}
\begin{equation}
\mathcal{L}_{l+1}^w=(l+1)(l+2)\left(B_r \frac{\mathrm{d}}{\mathrm{d}r}+\frac{B_r}{r}+\frac{\mathrm{d B_r}}{\mathrm{d}r} -(l+1)\frac{B_\theta}{r}\right).
\end{equation}
In above operators, we have used the following notations:
\begin{equation}
\omega_l=\omega+\frac{2m}{l(l+1)},
\end{equation}
\begin{equation}
l_p=l(l+1),
\end{equation}
\begin{equation}
q_l^m=\left(\frac{l^2-m^2}{4l^2-1}\right)^{1/2},
\end{equation}
\begin{equation}
\alpha_l^m=\frac{1}{l^2}q_l^m, \quad \beta_l^m=(l^2-1) q_l^m,
\end{equation}
and $B_r$ and $B_\theta$ represent the radial dependence of the background magnetic field $\bmath{B_0}$. For the dipolar field, we have
\begin{equation}
B_r=\frac{1}{r^3}, \quad B_\theta=\frac{1}{2r^3},
\end{equation}
while for the uniform vertical field
\begin{equation}
B_r=1, \quad B_\theta=-1.
\end{equation}
The vortical tidal forcing $\bmath f$ can be also project on to spherical harmonics
\begin{equation}\label{eq:fl}
f_{l}=-2\frac{\mathrm{i}m}{l(l+1)}\left(l(l+1)\frac{X_l(r)}{r}-r\frac{\mathrm{d}^2X_l(r)}{\mathrm{d}r^2}-2\frac{\mathrm{d}X_l(r)}{\mathrm{d}r}\right),
\end{equation}
\begin{equation}\label{eq:fl1}
f_{l-1}=-2\frac{q_l^m}{l}\left(\frac{\mathrm{d}X_l(r)}{\mathrm{d}r}+(l+1)\frac{X_l(r)}{r}\right),
\end{equation}
\begin{equation}\label{eq:fl2}
f_{l+1}=2\frac{q_{l+1}^m}{l+1}\left(\frac{\mathrm{d}X_l(r)}{\mathrm{d}r}-l\frac{X_l(r)}{r}\right).
\end{equation}
For a homogeneous fluid, $f_l\equiv0$ because $X_l(r)$ satisfies \citep[equation (86) in][]{Ogilvie2013MNRAS}
\begin{equation}
\frac{1}{r^2}\frac{\mathrm{d}}{\mathbb{d}r}\left(r^2\frac{\mathrm{d} X_l(r)}{\mathrm{d}r}\right)-\frac{l(l+1)}{r^2}X_l(r)=0.
\end{equation}

The boundary conditions can be also projected onto spherical harmonics. The stress-free boundary condition~(\ref{eq:stress_BC}) leads to 
\begin{equation}
u_l^m=\frac{\mathrm{d}}{\mathrm{d}r}\left(\frac{v_l^m}{r}\right)=\frac{\mathrm{d}}{\mathrm{d}r}\left(\frac{w_l^m}{r}\right)=0.
\end{equation}
The insulating boundary condition $(\grad\btimes\bmath{b})_r$ leads to $c_l^m=0$. The poloidal magnetic field continuously extends to the insulating regions as a potential field $\bmath B^e=-\grad P$, where the scalar potential $P$ satisfies Laplace's equation 
\begin{equation}
\nabla^2 P=0.
\end{equation} 
The solution of the above equation is 
\begin{equation}
P=\sum\sum g_l^m r^l Y_l^m(\theta,\phi),
\end{equation}
in the rigid inner core and
\begin{equation}
P=\sum\sum h_l^m r^{-(l+1)} Y_l^m(\theta,\phi),
\end{equation}
outside the body. Substituting those solutions into $\bmath B^e=-\grad P$ and comparing with the spherical harmonics expansion of $\bmath b$ in the fluid region, we obtain the boundary conditions
\begin{equation}\label{eq:ab_innerBC}
a_l^m(r)-lb_l^m(r)=0,
\end{equation} 
at the inner boundary and
\begin{equation}\label{eq:ab_outerBC}
a_l^m(r)+(l+1)b_l^m(r)=0,
\end{equation} 
at the outer boundary. Using the divergence-free condition of $\bmath b$ in equation~(\ref{eq:divblm0}), we can write the boundary conditions~(\ref{eq:ab_innerBC}-\ref{eq:ab_outerBC}) as
\begin{equation}
\frac{d a_l^m}{dr}-\frac{l-1}{r}a_l^m=0,
\end{equation}
at the inner boundary and 
\begin{equation}
\frac{d a_l^m}{dr}+\frac{l+2}{r}a_l^m=0,
\end{equation}
at the outer boundary.

\section{Frequency-averaged dissipation in the presence of a magnetic field}\label{App:FreAve}

\cite{Ogilvie2013MNRAS} found a way of calculating a certain frequency-average of the tidal response of a slowly and uniformly rotating barotropic fluid body to harmonic forcing. In this Appendix we consider how the argument and results presented in Section~4 of that paper are affected by the presence of a magnetic field. All the section numbers and equation numbers used below refer to \cite{Ogilvie2013MNRAS}.

The equilibrium condition~(21) is modified to
\begin{equation}
  0=-\grad(h+\Phi_\rmg+\Phi_\rmc)+\f{1}{\mu_0\rho}(\grad\btimes\bmB)\btimes\bmB
\end{equation}
and the linearized equations are modified to
\begin{equation}
  \ddot\bxi+2\bOmega\btimes\dot\bxi=-\grad W+\mathcal{F}\bxi,
\end{equation}
\begin{equation}
  W=h'+\Phi'+\Psi,
\end{equation}
\begin{equation}
  \rho'=-\grad\bcdot(\rho\bxi),
\end{equation}
\begin{equation}
  \nabla^2\Phi'=4\pi G\rho',
\end{equation}
where $\mathcal{F}$ is the linearized Lorentz force operator, which is self-adjoint with respect to a mass-weighted inner product and is given by
\begin{equation}
  \mathcal{F}\bxi=\f{1}{\mu_0\rho}\left[(\grad\btimes\bmB')\btimes\bmB+(\grad\btimes\bmB)\btimes\bmB'\right],
\end{equation}
where
\begin{equation}
  \bmB'=\grad\btimes(\bxi\btimes\bmB).
\end{equation}

In making the low-frequency asymptotic analysis in Section~4.4, we wish to assume that the frequencies of Alfv\'en and slow magnetoacoustic waves are, like those of the inertial waves, small compared to those of acoustic (or fast magnetoacoustic) and surface gravity waves. Unlike the inertial waves, however, the Alfv\'en and slow magnetoacoustic waves are not bounded in frequency if we allow ourselves to consider arbitrarily short wavelengths. We therefore need to apply a high-wavenumber cutoff to the response in order to contain the spectrum of low-frequency oscillations. This may be justified by assuming that the tidal response is smooth or by appealing to resistivity to eliminate disturbances of small scale.

The relevant scaling assumptions are then that the Lehnert number is $O(1)$ (or smaller) and that the perturbations are of large scale. Formally this can be achieved by saying that both $\Omega$ and $\bmB$ are $O(\epsilon)$. With the arbitrary normalization $\Psi=O(1)$, we then have (as before) $\bxi=O(1)$, $W=O(\epsilon^2)$, $h'=O(1)$, $\Phi'=O(1)$ and now $\bmB'=O(\epsilon)$. Our reduced system of linearized equations at leading order is then
\begin{equation}
  \ddot\bxi+2\bOmega\btimes\dot\bxi=-\grad W+\mathcal{F}\bxi,
\end{equation}
\begin{equation}
  h'+\Phi'+\Psi=0,
\end{equation}
\begin{equation}
  \rho'=-\grad\bcdot(\rho\bxi),
\end{equation}
\begin{equation}
  \nabla^2\Phi'=4\pi G\rho',
\end{equation}
and is to be solved on a spherically symmetric, hydrostatic basic state unaffected by rotation or magnetic fields.

We again decompose the perturbations into non-wavelike and wavelike parts, satisfying respectively
\begin{equation}
  \ddot\bxi_\rmnw=-\grad W_\rmnw,
\end{equation}
\begin{equation}
  h'_\rmnw+\Phi'_\rmnw+\Psi=0,
\end{equation}
\begin{equation}
  \rho'_\rmnw=-\grad\bcdot(\rho\bxi_\rmnw),
\end{equation}
\begin{equation}
  \nabla^2\Phi'_\rmnw=4\pi G\rho'_\rmnw,
\end{equation}
and
\begin{equation}
  \ddot\bxi_\rmw+2\bOmega\btimes\dot\bxi_\rmw=-\grad W_\rmw+\mathcal{F}\bxi_\rmw+\bmf,
\label{xiwdd}
\end{equation}
\begin{equation}
  \grad\bcdot(\rho\bxi_\rmw)=0,
\label{anelastic}
\end{equation}
where $\rho'_\rmw=h'_\rmw=\Phi'_\rmw=0$ and
\begin{equation}
  \bmf=-2\bOmega\btimes\dot\bxi_\rmnw+\mathcal{F}\bxi_\rmnw
\end{equation}
is the effective force per unit mass driving the wavelike part of the solution. As before, the non-wavelike tide may be assumed to be instantaneously related to the tidal potential through
\begin{equation}
  \bxi_\rmnw=-\grad X,
\end{equation}
where $X$ is the solution of the elliptic equation~(61). The energy equation for the wavelike part is
\begin{equation}
  \f{\rmd}{\rmd t}\left[\f{1}{2}\int\rho\left(|\bmu_\rmw|^2-\bxi_\rmw\bcdot\mathcal{F}\bxi_\rmw\right)\rmd V\right]=\int\rho\bmu_\rmw\bcdot\bmf\,\rmd V,
\end{equation}
where $\bmu_\rmw=\dot\bxi_\rmw$ is the wavelike velocity and the second term in the integral on the left-hand side is the magnetic energy associated with the wavelike displacement.

Turning now to the impulsive forcing analysed in Section~4.6, we again consider a tidal potential of the form
\begin{equation}
  \Psi=\hat\Psi(\bmr)H(t),
\end{equation}
where $H(t)$ is the Heaviside step function. This implies that
\begin{equation}
  \bxi_\rmnw=\hat\bxi_\rmnw(\bmr)H(t),
\end{equation}
leading to an effective force
\begin{equation}
  \bmf=\hat\bmf(\bmr)\delta(t)+\tilde\bmf(\bmr)H(t),
\end{equation}
where $\delta(t)$ is the Dirac delta function,
\begin{equation}
  \hat\bmf=-2\bOmega\btimes\hat\bxi_\rmnw
\end{equation}
derives solely from the Coriolis force and
\begin{equation}
  \tilde\bmf=\mathcal{F}\hat\bxi_\rmnw
\end{equation}
derives solely from the Lorentz force. Therefore the impulsive contibution to the effective force comes only from the Coriolis force and not from the Lorentz force. The solution of equations (\ref{xiwdd}) and (\ref{anelastic}) in this case involves a wavelike displacement $\bxi_\rmw$ that is continuous in $t$ but has a discontinuous first derivative at $t=0$. The wavelike velocity immediately after the impulse is again
\begin{equation}
  \hat\bmu_\rmw=\hat\bmf-\grad\hat W_\rmw,
\end{equation}
where $\hat W_\rmw$ is chosen to satisfy the anelastic constraint $\grad\bcdot(\rho\hat\bmu_\rmw)=0$  and the boundary conditions $\hat u_{\rmw,r}=0$. The energy transferred in the impulse is equal to the kinetic energy immediately after the event,
\begin{equation}
  \hat E=\f{1}{2}\int\rho|\hat\bmu_\rmw|^2\,\rmd V.
\end{equation}
There is no change in the wavelike magnetic energy at $t=0$ because $\bxi_\rmw$ is continuous there. Since $\hat\bmu_\rmw$ 
derives solely from the Coriolis force, we conclude that the magnetic field has no effect either on the impulsive energy transfer associated with this term, or on the frequency-averaged dissipation related to it through equation 100.  We note, however, that the Lorentz part of the effective force in equation (B23), which is neglected in the main part of this paper, could alter the energy of the wave-like disturbance after the impulse at $t=0$ and therefore make an additional contribution to the frequency-averaged dissipation, which requires further investigation.


%%%%%%%%%%%%%%%%%%%%%%%%%%%%%%%%%%%%%%%%%%%%%%%%%%

% Don't change these lines
\bsp	% typesetting comment
\label{lastpage}
\end{document}